\documentclass[twocolumn]{aastex63}
\usepackage[title]{appendix}
\usepackage{graphicx}
\usepackage{amssymb}
\usepackage{amsmath}
\usepackage{color}
\usepackage{url}
\accepted{Oct 19, 2019, AJ}

\let\pwiflocal=\iffalse \let\pwifjournal=\iffalse
\usepackage[utf8]{inputenc}
\usepackage{textcomp}
\usepackage{amsmath,amssymb,gensymb,times,graphicx,morefloats,color}
\usepackage{afterpage, bm}
\usepackage{natbib,hyperref}
\usepackage[outdir=./]{epstopdf}
\usepackage{mathrsfs}
\newcommand{\hal}{H$\alpha$}
\newcommand{\rsun}{$R_\odot$}

\newcommand{\msun}{$\rm{M}_\odot$}

\newcommand{\degs}{$^\circ$}

\newcommand{\vsini}{$v$sin$i$}

\newcommand{\mdot}{$\dot{\rm{M}}$}
\newcommand{\hei}{He {\scshape i}}

\newcommand{\oi}{[O {\scshape i}]}
\newcommand{\caii}{Ca {\scshape ii}}

\newcommand{\ari}{Ar {\scshape i}}
\newcommand{\kms}{km s$^{-1}$}
\newcommand{\gaia}{{\it Gaia}}

\shortauthors{Tofflemire et al.}
\shorttitle{Accretion Kinematics in TWA 3A}
\begin{document}
\correspondingauthor{Benjamin M. Tofflemire}
\email{tofflemire@utexas.edu}

\title{Accretion Kinematics in the T Tauri Binary TWA 3A: \\ Evidence for Preferential Accretion onto the TWA 3A Primary}

\author[0000-0003-2053-0749]{Benjamin M. Tofflemire}
\affiliation{Department of Astronomy, The University of Texas at Austin, 2515 Speedway, Stop C1400, Austin, TX 78712, USA}
\affiliation{Department of Astronomy, University of Wisconsin--Madison,
  475 North Charter Street, Madison, WI 53706, USA}
  
\author{Robert D. Mathieu}
\affiliation{Department of Astronomy, University of Wisconsin--Madison,
  475 North Charter Street, Madison, WI 53706, USA}

\author{Christopher M. Johns-Krull}
\affiliation{Physics \& Astronomy Department, Rice University, 6100 Main St., Houston, TX 77005, USA}

\begin{abstract}

We present time-series, high-resolution optical spectroscopy of the eccentric T Tauri binary TWA 3A. Our analysis focuses on variability in the strength and structure of the accretion tracing emission lines \hal \ and \hei \ 5876\AA. We find emission line strengths to display the same orbital-phase dependent behavior found with time-series photometry, namely, bursts of accretion near periastron passages. Such bursts are in good agreement with numerical simulations of young eccentric binaries. During accretion bursts, the emission of \hei \ 5876\AA \ consistently traces the velocity of the primary star. After removing a model for the system's chromospheric emission, we find the primary star typically emits $\sim$70\% of the \hei \ accretion flux. We interpret this result as evidence for circumbinary accretion streams
that preferentially feed the TWA 3A primary. This finding is in contrast to most numerical simulations, which predict the secondary should be the dominant accretor in a binary system. Our results may be consistent with a model in which the precession of an eccentric circumbinary disk gap alternates between preferentially supplying mass to the primary and secondary. 

\end{abstract}

\keywords{stars: individual (TWA 3A) --- stars: formation --- binaries:
  close --- accretion, accretion disks}

\section{INTRODUCTION}

\label{intro}
Protoplanetary disks are integral to the formation of low-mass stars and planets. In the case of single stars, a well-established paradigm has been developed describing the star-disk interaction \citep{Koenigl1991,Hartmannetal1994}. In this model, a star's magnetic field truncates the inner regions of the disk, channeling flows of material along magnetic field lines before impacting the stellar surface at their foot points \citep{Shuetal1994}. This magnetic accretion scenario has been able to describe many of the observational characteristics of young accreting systems, commonly refereed to as classical T Tauri stars (CTTSs). 
Much of the theory for stellar evolution in the pre-main-sequence (PMS) phase (e.g., angular momentum evolution; \citealt{Krishnamurthietal1997}) and the evolution of disk material (e.g., photoevaporation; \citealt{Pascucci&Sterzik2009,ErcolanoPascucci2017}) are underpinned by this theory of the star-disk interaction \citep{Hartmannetal2016}.

Binary systems, however, are a common outcome of star formation \citep{Raghavanetal2010,Krausetal2011} that can fundamentally alter the canonical star-disk interaction. Departures from the single-star interaction are most significant in cases of short-period systems where orbital resonances are capable of dynamically clearing the central region of disk material on the size scale of the binary orbit \citep[2--3 times the binary semi-major axis $a$;][]{Artymowicz&Lubow1994}. In this disk architecture, steady inward flows of disk material are predicted to give way to periodic streams that bridge the gap between a circumbinary disk and the central stars \citep{Artymowicz&Lubow1996,Gunther&Kley2002,DeVal-Borroetal2011}. 

For binaries with circular orbits, these streams are predicted to occur on the orbital timescale of material at the inner edge of the circumbinary disk, where a buildup of material becomes unstable and falls into the cleared region. These inward flows are predicted to periodically enhance stellar accretion rates on a timescale longer than the binary orbit (e.g., \citealt{Farrisetal2014}). 

In the presence of even slight orbital eccentricities, circumbinary streams are predicted every binary orbital period as apastron passages draw in material from the inner edge of the circumbinary disk \citep{Mirandaetal2017}. These streams fuel bursts of accretion whose timing and amplitude depend on the binary orbital parameters. In the case of highly eccentric binaries, like TWA 3A (the subject of this study), accretion bursts are predicted near each periastron passage that increase the specific accretion rate by a factor of $\sim$10 \citep{Munoz&Lai2016}. Evidence for such orbit-modulated accretion has been shown in the highly eccentric binaries DQ Tau ($e$=0.57; \citealt{Mathieuetal1997,Tofflemireetal2017a,Muzerolleetal2019}) and TWA 3A ($e$=0.63; \citealt{Tofflemireetal2017b}) through long-term photometric monitoring. Additionally, DQ Tau has also shown periodic increases in the level of absorption line veiling and \hal \ luminosity near periastron \citep{Basrietal1997}. In the lower eccentricity system UZ Tau E ($e$=0.33; \citealt{Jensenetal2007}), we note that accretion appears more chaotic and not directly tied to the binary orbital period.

While the observations of TWA 3A and DQ Tau provide a fair match to numerical simulations, they bring into question the exact way in which material reaches the stellar surface. In high-resolution hydrodynamic simulations, each star develops a small circumstellar disk that collects incoming stream material. Periastron accretion bursts in this scenario are driven more by tidal torques that each star exerts on its companion's disk rather than from the introduction of new material to their circumstellar disks via streams \citep{Munoz&Lai2016}. Short-period systems ($P \leq 35$ d), however, have close periastron passages ($\sim$10 stellar radii) and when considering the disruptive effect stellar magnetic fields have on inner disk material (assumed to truncate material at $\sim$5$R_\star$; \citealt{Johnstoneetal2014}), it becomes unlikely that either star hosts a stable circumstellar disk. Under these binary orbital parameters it is uncertain how stream material interacts with the stars and their magnetic fields. No simulations to date have ventured into this domain. 

To gain further insight into accretion processes in short-period binaries, we have monitored the eccentric, T Tauri binary TWA 3A with high-resolution optical spectroscopy. These observations overlap with the long-term, moderate-cadence photometry presented in \citet{Tofflemireetal2017b}. Our observations focus on $\sim$3 orbital periods in which we seek to characterize the kinematic properties of accretion during the rise of periastron accretion bursts. We achieve this through the analysis of accretion-tracing emission lines that probe the foot-points of magnetic accretion flows. We also investigate the emission of outflow-tracing lines with respect to variable accretion and orbital motion. These observations, coupled with the wealth of information known about this system, make it a unique setting to constrain the dynamical properties of binary accretion. 

\

\section{The TWA 3 System}
\label{twa3a}

In this section we describe the stellar orbits and distribution of disk material in the young, hierarchical triple system TWA 3. Much of following is a summary of the study by \citet{Kelloggetal2017}, where we highlight their results for the inner binary, TWA 3A. 

TWA 3 is a member of the $\sim$10 Myr TW Hya association \citep{Mamajek2005}, composed of an inner spectroscopic binary, TWA 3A, and an outer triple companion, TWA 3B, separated by $1\farcs55$. \gaia \ measurements \citep{GAIA2016,GAIAdr2} for the A and B components suffer from significant excess astrometric noise, which is common to sources at this magnitude ($G\sim11$) and may be impacted by photometric variability \citep{Lindegrenetal2018}. We hesitate to over-interpret these data, but note that the parallax distances are consistent for each source, when accounting for systematic noise (following the Bayesian distance estimation in \citealt{ABJ2016}), and generally agrees with the TWA 3A--3B orbit arcs fit in \citet{Kelloggetal2017}. Future \gaia \ data releases will improve these measurements but for the current work, the large separation means that the influence of TWA 3B on the inner binary and its disk material is negligible over the time scales we consider here. 

\begin{figure*}[!t]
  \centering
  \includegraphics[keepaspectratio=true,scale=1.05]{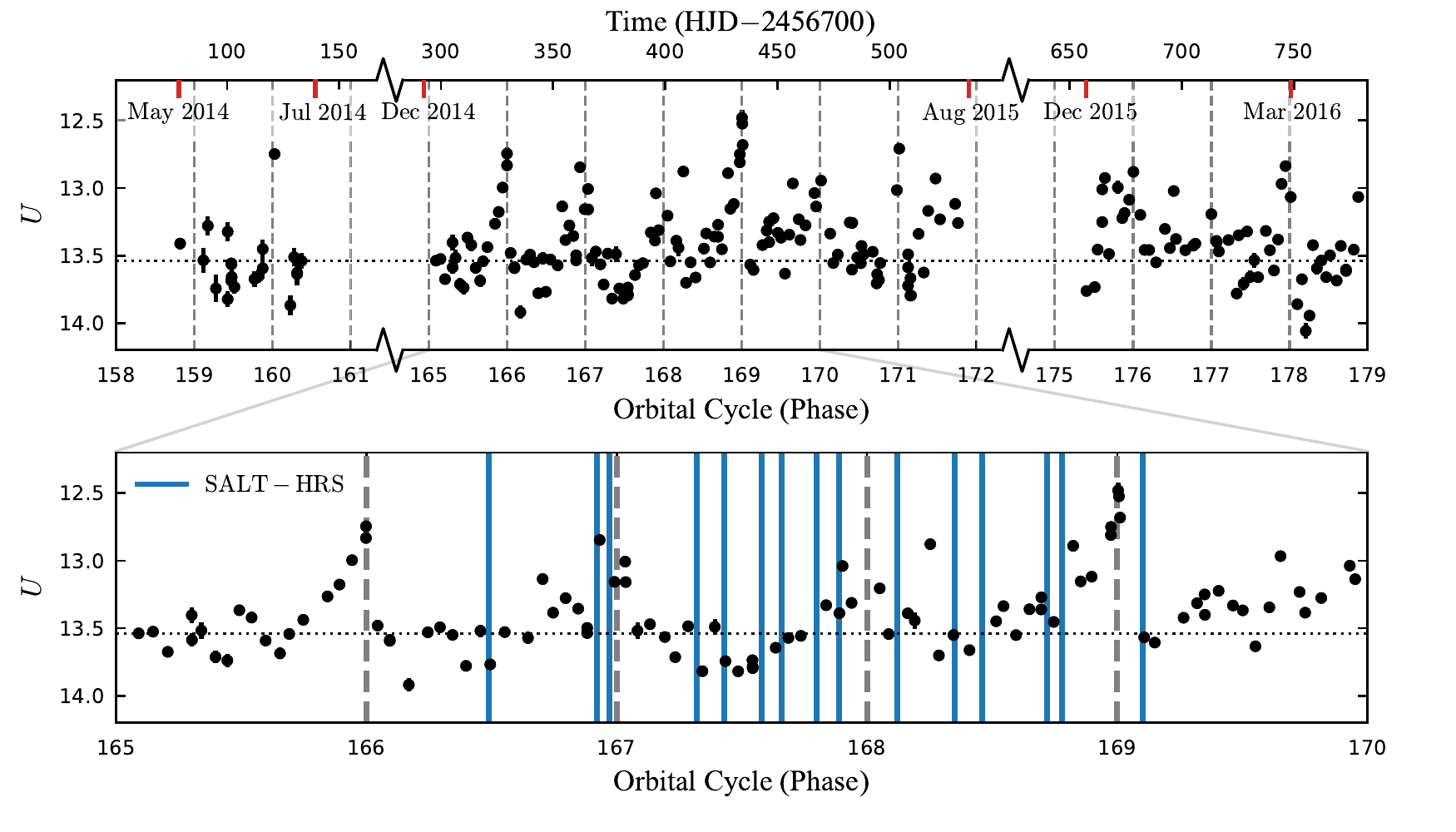}
  \caption{Overview of our LCO-SALT time-series monitoring campaign of TWA 3. {\bf Top:} TWA 3 combined $U$-band light curve as a function of TWA 3A orbital cycle on the bottom axis and heliocentric Julian date on the top axis. Vertical dashed lines mark TWA 3A periastron passages and the horizontal dotted line marks the quiescent $U$-band luminosity (average of orbital phases $\phi$ = 0.2--0.4). These data were originally published in \citet{Tofflemireetal2017b}. {\bf Bottom:} A zoom-in of five TWA 3A orbital periods where SALT--HRS observations are denoted with vertical blue bands.}
  \label{fig:Ulc}
\end{figure*}

The inner binary, TWA 3A, hosts an extended circumbinary disk with signatures of active accretion \citep{Herczegetal2009,Tofflemireetal2017b}. ( A recent non-detection of accretion activity in TWA 3A by \citet{Venutietal2019} occurred at an orbital phase of $\phi$=0.23, corresponding to the periodic accretion minimum (see Figure \ref{fig:Ulc}, also \citealt{Tofflemireetal2017b})). TWA 3B on the other hand, shows no evidence for disk material \citep{Jayawardhanaetal1999,Andrewsetal2010} and has the spectral properties of a non-accreting, weak-lined T Tauri star \citep{Herczeg&Hillenbrand2014,Venutietal2019}. Table \ref{tab:TWA} presents the TWA 3A spectroscopic orbital solution and properties of the circumbinary disk.

\begin{deluxetable}{lcc}
\tablewidth{0pt}
\tabletypesize{\footnotesize}
\tablecaption{TWA 3A System Characteristics}
\tablehead{
  \colhead{Parameter} &
  \colhead{Value} &
  \colhead{References}}
\startdata
P (days)                         & $34.87846 \pm 0.00090$ & (1) \\
$\gamma$ (km s$^{-1}$)           & $+10.17 \pm 0.40$      & (1) \\
$e$                              & $0.6280 \pm 0.0060$    & (1) \\
$\omega$  ($^\circ$)              & $80.5 \pm 1.2$         & (1) \\
T$_{\rm peri}$ (HJD-2,400,000)   & $51239.659 \pm 0.063$& (1)$^a$\\
$a$ sin $i$ ($R_\odot$)           & $27.34 \pm 0.29$     & (1) \\
$q\equiv M_2/M_1$                & $0.841 \pm 0.014$      & (1) \\
$M_1$ sin$^3$ $i$ ($M_\odot$)     & $0.1224 \pm 0.0055$    & (1) \\
$M_2$ sin$^3$ $i$ ($M_\odot$)     & $0.1030 \pm 0.0043$    & (1) \\
$v$ sin $i$ (km s$^{-1}$)        & $<8$                    & (1)$^b$ \\
$i_{\rm binary}$ ($^\circ$)         & $\sim35-53$    & (1)$^c$ \\
$i_{\rm disk}$ ($^\circ$)           & $\sim36$           & (2)$^c$ \\
Disk $M_{\rm dust}$ ($M_\odot$)    & $7\times10^{-6}$        & (2) \\
d (pc)                           & $\sim36.6$       & (3)\\
$A_V$                            & $0.04 \pm 0.3$          & (4) 
\enddata 
\tablerefs{(1) \citet{Kelloggetal2017}, (2) \citet{Andrewsetal2010}, (3) This work, following \citet{ABJ2016} -- no error estimate is given due to the
source’s excess astrometic error, (4) \citet{Tofflemireetal2017b}}
\tablenotetext{a}{The T$_{\rm peri}$ value presented is 42 orbital periods prior to the value in \citet{Kelloggetal2017}, placing the earliest archival spectra at an orbital cycle of zero. The error is assumed to be the same.} 
\tablenotetext{b}{Upper limit for both stars, below the instrumental resolution element.}
\tablenotetext{c}{$i$ has a 180\degs \ ambiguity from unknown line of ascending nodes.}
\label{tab:TWA}
\end{deluxetable}

The distribution of stars and disk material in this system suggests that multiplicity has influenced the system's evolution. The TWA 3A circumbinary disk has been modeled as having both inner and outer truncations \citep{Andrewsetal2010}. The inner truncation at $\sim$0.3 AU is presumably formed by binary orbital resonances and agrees fairly well with the predicted inner disk gap size of 2--3 $a$ \citep{Artymowicz&Lubow1994}. The outer truncation at a projected radius of $\sim$20 AU is likely to have been shaped by the triple companion's orbital motion. Even without a well-constrained TWA 3A-3B orbit, the TWA 3A disk radius to companion separation falls within the sample of Taurus binaries observed by \citet{Manaraetal2019} with ALMA, suggesting dynamical truncation. It is additionally unclear what processes led to circumstellar material surrounding one component of the system, TWA 3A, while not the other.  

Furthermore, there is evidence that the TWA 3A binary orbital plane may be misaligned from the large-scale circumbinary disk inclination observed by \citet{Andrewsetal2010}. A spatially resolved observation of TWA 3Aa--3Ab with VLTI/PIONIER \citep{Anthoniozetal2015}, coupled with the spectroscopic orbital parallax, results in two families of allowed binary inclinations. The first family has small mutual binary-disk inclinations, $\psi$, which include coplanarity ($\psi$ $<$ 20\degs), while the second includes large mutual inclinations ($\psi$ $\sim$ 90\degs). 

Additional support for a misalignment between the binary and circumbinary disk inclination comes from the mismatch between the dynamical masses assuming disk-binary coplanarity ($M_1$=0.6 \msun; $M_2$=0.5 \msun) and the observed luminosities and spectral types (M3--M4.5), which would correspond to $\sim$0.3 \msun \ stars \citep{Herczeg&Hillenbrand2014}. Even with known systematic offsets in measurements of PMS spectral types (see \citealt{Gully-Santiagoetal2017} and references therein), these measurements disagree considerably. Notably though, both families of possible binary orbital inclinations include those that eliminate this tension. 

As this study is aimed at characterizing the interaction between short-period binaries and their inner disk material, the mutual inclination of the binary and disk orbital planes is particularly relevant. Analytic theory and numerical simulations of misaligned binary-disk orbital planes have been developed recently to explain the growing observational evidence for binary-disk misalignment (e.g., KH 15D - \citealt{Capeloetal2012}; GG Tau - \citealt{Cazzoletti2017}; 99 Her - \citealt{Kennedyetal2012}). In the low-misalignment regime, \citet{Juhasz&Facchini2017} find steady-state disk warps that agree with linear wave-propagation theory for misalignments $\psi$$<$30\degs \ \citep{Papaloizou&Lin1995,Lubow&Ogilvie2000}. In this scenario, the inner part of the circumbinary disk is coplanar with the binary orbit.

For larger misalignments, \citet{Martin&Lubow2017} find that low-mass disks around high-eccentricity binaries can quickly become polar given a large initial misalignment ($\psi$ $\gtrsim$ 40\degs). The result of this process leaves the circumbinary disk's angular momentum vector perpendicular to the binary's, i.e.\ the apastron passage is directly out of the plane of the circumbinary disk (see \citealt{Lubow&Martin2018}). And indeed, a binary with this configuration has recently been discovered \citep[HD 98800;][]{Kenedyetal2019}.

Comparing these two possible scenarios in the light of the time-series photometry in \citet{Tofflemireetal2017b}, we find 
that the agreement between the observed accretion variability to models of coplanar binary accretion, and also to DQ Tau (a binary with known binary-disk alignment; \citealt{Czekalaetal2016}), supports binary orbital inclinations with a modest mutual disk inclination (i.e,.\ a coplanar inner circumbinary disk). 

For the remainder of this work we assume a binary inclination of $i_{\rm{binary}}$=44\degs \ (average of the low-mutual inclination solutions) and adopt the following system parameters: stellar masses of 0.37 \msun \ and 0.31 \msun, and a semi-major axis of 39.3 \rsun with periastron and apastron separations of 14.7 and 64.0 \rsun, respectively. We set the radius of the primary and secondary stars to 0.78 and 0.72 \rsun, respectively, from the average of \citet{Dotteretal2008} stellar evolution models with ages between 5--20 Myr at the corresponding masses. (These adopted parameters and \gaia \ DR2 updated distance slightly increase the mass accretion rate values presented in \citet{Tofflemireetal2017b}, which assumed binary-disk coplanarity.)

\section{OBSERVATIONS \& DATA REDUCTION}
\label{obsdr}

From 2014 May to 2016 April, TWA 3 was the target of a multi-site, time-series monitoring campaign combining multi-color optical photometry and high-resolution optical spectroscopy. To show the relative sequencing of these two campaigns, Figure \ref{fig:Ulc} presents the $U$-band light curve obtained from the Las Cumbres Observatory (LCO) 1m Network with spectroscopic observations from the Southern African Large Telescope (SALT) High Resolution Spectrograph (HRS) marked with vertical blue bands. Vertical dashed lines denote TWA 3A periastron passages.

The following subsections describe the observations and data reduction methods for SALT--HRS spectroscopy and LCO imaging. A description of archival optical spectra that are also presented in this paper is included below.

\begin{deluxetable*}{lccrrrrr}
\tablewidth{0pt}
\tabletypesize{\footnotesize}
\tablecaption{Date, Accretion Rate, Veiling, and Equivalent Widths for TWA 3A SALT--HRS
  Observations} 
\tablehead{
  \colhead{HJD} &
  \colhead{Orbital} &
  \colhead{\mdot} &
  \colhead{Veiling} &
  \colhead{H $\alpha$} &
  \colhead{H $\beta$} &
  \colhead{\hei \ 5876\AA} &
  \colhead{\oi \  6300\AA} \\
  \colhead{} &
  \colhead{Phase} &
  \colhead{($10^{-11}$\msun \ yr$^{-1}$)$^a$} &
  \colhead{} &
  \colhead{(\AA)$^b$} &
  \colhead{(\AA)$^b$} &
  \colhead{(\AA)$^b$} &
  \colhead{(\AA)$^c$} 
}
\startdata
2457046.4476 & 166.49 & $ 5.8\pm 0.6$ & $0.04\pm0.01$ & $-11.56\pm0.82$ & $-6.43\pm0.20$ & $-0.63\pm0.06$  & $-0.40\pm0.06 $ \\ 
2457061.4090 & 166.92 & $23.8\pm 0.9$ & $0.15\pm0.02$ & $-33.33\pm3.09$ & $-14.68\pm1.02$ & $-1.63\pm0.16$ & $-0.39\pm0.05 $ \\ 
2457063.3984 & 166.97 & $21.8\pm 0.8$ & $0.13\pm0.02$ & $-49.34\pm4.96$ & $-15.64\pm1.17$ & $-1.08\pm0.10$ & $-0.40\pm0.05 $ \\ 
2457075.3713 & 167.32 & $ 6.7\pm 0.6$ & $0.06\pm0.01$ & $-11.32\pm0.75$ & $-8.91\pm0.46$ & $-0.71\pm0.07$  & $-0.43\pm0.06 $ \\ 
2457079.3610 & 167.43 & $ 6.1\pm 0.6$ & $0.03\pm0.01$ & $-16.60\pm1.43$ & $-8.94\pm0.51$ & $-0.87\pm0.10$  & $-0.45\pm0.06 $ \\ 
2457084.5985 & 167.58 & $ 5.7\pm 0.5$ & $0.04\pm0.01$ & $-17.52\pm1.52$ & $-7.29\pm0.31$ & $-0.81\pm0.09$  & $-0.47\pm0.07 $ \\ 
2457087.3351 & 167.66 & $ 7.8\pm 0.4$ & $0.07\pm0.01$ & $-12.14\pm0.84$ & $-7.25\pm0.26$ & $-0.85\pm0.08$  & $-0.41\pm0.06 $ \\ 
2457092.3205 & 167.80 & $11.7\pm 0.6$ & $0.07\pm0.01$ & $-19.83\pm1.72$ & $-10.91\pm0.68$ & $-0.90\pm0.08$ & $-0.52\pm0.07 $ \\ 
2457095.3155 & 167.89 & $12.2\pm 0.9$ & $0.09\pm0.02$ & $-25.66\pm2.41$ & $-13.14\pm0.95$ & $-1.82\pm0.19$ & $-0.61\pm0.08 $ \\ 
2457103.5442 & 168.12 & $10.3\pm 0.6$ & $0.08\pm0.02$ & $-11.93\pm0.83$ & $-9.79\pm0.58$ & $-0.92\pm0.08$  & $-0.48\pm0.06 $ \\ 
2457111.5225 & 168.35 & $ 8.7\pm 0.5$ & $0.08\pm0.02$ & $-15.23\pm1.22$ & $-10.98\pm0.72$ & $-1.10\pm0.11$ & $-0.58\pm0.08 $ \\ 
2457115.2590 & 168.46 & $ 8.7\pm 0.5$ & $0.03\pm0.01$ & $-11.55\pm0.84$ & $-6.03\pm0.22$ & $-0.77\pm0.08$  & $-0.28\pm0.05 $ \\ 
2457124.4924 & 168.72 & $12.9\pm 0.5$ & $0.08\pm0.02$ & $-27.01\pm2.60$ & $-14.08\pm1.08$ & $-1.15\pm0.11$ & $-0.65\pm0.08 $ \\ 
2457126.4848 & 168.78 & $17.8\pm 0.6$ & $0.08\pm0.02$ & $-39.76\pm4.10$ & $-18.57\pm1.60$ & $-1.66\pm0.18$ & $-0.61\pm0.08 $ \\ 
2457137.4515 & 169.10 & $10.5\pm 0.5$ & $0.08\pm0.02$ & $-20.75\pm1.87$ & $-15.06\pm1.19$ & $-1.23\pm0.12$ & $-0.55\pm0.07 $ \\ 
\enddata 
\label{tab:obs}
\tablenotetext{a}{Mass accretion rate interpolate from adjacent photometric
  measurements.}
\tablenotetext{b}{Equivalent width measurements corrected for triple companion using Equation \ref{eqn:ew}}
\tablenotetext{c}{Equivalent width measurements corrected for triple companion using Equation \ref{eqn:ew_oi}}
\end{deluxetable*}

\begin{deluxetable}{lcrrr}
\tablewidth{0pt}
\tabletypesize{\footnotesize}
\tablecaption{Date, Veiling, and Emission Line Equivalent Widths for TWA 3A FEROS Observations}
\tablehead{
  \colhead{HJD} &
  \colhead{Orbital} &
  \colhead{Veiling} &
  \colhead{H$\alpha$} &
  \colhead{\hei \ 5876\AA} \\
  \colhead{}&
  \colhead{Phase}&
  \colhead{}&
  \colhead{(\AA)$^a$}&
  \colhead{(\AA)$^a$}
}
\startdata
2451260.5221 & 0.6 & $0.02\pm0.01$ & $-23.33\pm3.77$ & $-1.83\pm0.33 $ \\ 
2451331.5915 & 2.64 & $0.04\pm0.01$ & $-15.93\pm2.19$ & $-1.20\pm0.20 $ \\ 
2451621.5378 & 10.95 & $0.14\pm0.02$ & $-94.02\pm16.28$ & $-3.53\pm0.61 $ \\ 
2451622.5731 & 10.98 & $0.2\pm0.02$ & $-100.31\pm16.47$ & $-3.28\pm0.53 $ \\ 
2451623.6000 & 11.01 & $0.09\pm0.01$ & $-44.40\pm7.42$ & $-3.60\pm0.64 $ \\ 
2451624.5913 & 11.04 & $0.07\pm0.02$ & $-25.77\pm4.07$ & $-1.61\pm0.28 $ \\ 
2451625.5186 & 11.06 & $0.13\pm0.02$ & $-19.91\pm2.65$ & $-1.42\pm0.22 $ \\ 
2451733.4665 & 14.16 & $0.06\pm0.02$ & $-17.02\pm2.39$ & $-0.74\pm0.12 $ \\ 
2451737.5133 & 14.27 & $0.1\pm0.03$ & $-21.98\pm3.27$ & $-1.09\pm0.17 $ \\ 
\enddata 
\label{tab:feros}
\tablenotetext{a}{Equivalent width measurements corrected for triple companion using Equation \ref{eqn:ew}.}
\end{deluxetable}

\subsection{SALT--HRS}
\label{salt}

Located in Sutherland, South Africa, SALT is a fixed-elevation telescope with a 10-m class primary mirror \citep{Buckleyetal2006}. Observations of TWA 3 were made with the dual-beam, fiber-fed, echelle High-Resolution Spectrograph (HRS; \citealt{Bramalletal2010,Bramalletal2012}). 
 In the medium-resolution mode, a 500 $\mu$m ($2\farcs2$) fiber is image-sliced into a vertical pseudo-slit providing spectral resolutions up to $R$ $\sim$ 40,000 across a broad optical wavelength range in separate blue (3700-5500\AA) and red (5500-8900\AA) arms.

Critically for this study, SALT is queue scheduled, allowing for observations of specific TWA 3A orbital phases spread over multiple orbital periods. Our observing program aimed to detect the rise of the accretion rate towards periastron by making targeted observations near apastron passages and at orbital phases leading up to periastron passage. Over 107 days, 15 spectroscopic observations were made spanning $\sim$3 TWA 3A orbital periods. Each visit consisted of three $\sim$12 minute integrations with some variability due to the SALT track length. Table \ref{tab:obs} presents the heliocentric Julian date of our SALT observations, their corresponding orbital phase, relevant spectroscopic measurements, and the mass accretion rate  at the time of the observation (determined from an interpolation of the nearest mass-accretion rate measurements derived from $U$-band photometry, see Section \ref{lcogt}).

The reduction of HRS spectroscopy was performed with standard IRAF\footnote{IRAF is distributed by the National Optical Astronomy Observatory, which is operated by the Association of Universities for Research in Astronomy (AURA) under a cooperative agreement with the National Science Foundation.} tasks within a PyRAF script. Reduction steps include basic CCD calibrations (gain correction, over scan subtraction, bias subtraction, and image trimming), spectral flat correction, cosmic ray rejection \citep{vanDokkum2001}, and a scattered light correction. The three target spectra per visit are then extracted and dispersion corrected by a nightly ThAr comparison spectrum before being median combined. Red-arm spectral orders are then combined, weighted by the blaze function. 

In the blue arm, many orders do not detect the stellar continuum, or only detect it at the peak of the order's sensitivity. Combining orders becomes problematic in this case where the relative height of emission lines to the continuum (or noise level) is different in adjacent orders. For instance, an emission line in a low-sensitivity region of one order with no detected stellar continuum will have a relative height that is less than the same line in the adjacent order where continuum is detected. Combining these orders would artificially reduce the relative height and EW of the line. For this reason, blue-arm orders are not combined and the analysis of spectral lines is done in the order where the line has the highest sensitivity.

We note that with SALT's median seeing of $1\farcs3$ and HRS's $2\farcs2$ fiber, our spectra contain light from TWA 3A and its tertiary companion TWA 3B. In Section \ref{analysis}, we describe our spectral decomposition and procedure to remove the triple's contribution. 

\subsubsection{HRS Red Arm-Specific Reductions}
\label{redarm}

In the SALT--HRS red arm, two separate distortions of the slit image are present that affect the width of the spectral resolution element across a given order. The first is that the width of the imaged slit is wider (larger FWHM) on the long-wavelength side of an order. This effect is minor and we do not correct for it. 

The second is that the slit image is tilted by an angle that increases with pixel column (increasing toward the long-wavelength edge of the chip). This variable tilt causes significant spectral smearing that increases across the spectral order. This becomes problematic when combining adjacent orders where the resultant line shapes would then depend on the combination of two different spectral resolutions. Sky lines, and their removal are particularly affected, being the width of the resolution element.

In Appendix \ref{ap:red} we provide a full description of our method for correcting this distortion. In short, we use the emission lines in Th-Ar comparison images to measure the angle of the slit tilt across spectral orders. With multiple lines spanning the spectral order, we map the slit tilt (represented by an emission line's peak across the vertical extent of the order) as a function of the pixel column (dispersion axis) and pixel row (slit axis). For each order, a function is fit to the slit location in each row that is used to perform a transformation along the pixel columns (dispersion axis). This amounts to a stretch or a compression along an individual rows that aligns the slit image to its value at the order's center. 

We note that this process involves interpolation which introduces covariance between pixels. This may be relevant in the determination of equivalent width (EW) uncertainties, which is discussed in Section \ref{ew_meas_corr}. Flux conservation can also be a concern during interpolation, but with measures described in Appendix \ref{ap:red}, flux values are preserved to better than 1\% in these transformations. 

\subsection{Archival Spectroscopic Observations}
\label{archival}

To determine whether the trends observed over the $\sim$3 orbital periods monitored here are long lasting, we also compile archival observations of TWA 3, which are presented in \citet{Kelloggetal2017}. Specifically, we make use of FEROS spectra, which have similar wavelength coverage (3600 to 9200\AA), spectral resolution ($R$$\sim$44,000), and orbital phase coverage to our SALT--HRS observations. 

FEROS is a fiber-fed echelle spectrograph which was located on the ESO 1.5m telescope in La Silla, Chile during the time of the observations. Between 1999 March and 2000 July, 9 spectra were obtained spanning $\sim$15 orbital cycles. Table \ref{tab:feros} provides a list of the FEROS observation dates, their orbital cycles and phases, and select spectroscopic measurements. For a full description of these data we refer readers to \citet{Kelloggetal2017}. We note that due to the FEROS 2$\arcsec$ fiber, these spectra also contain flux from TWA 3B. 

\subsection{LCO 1m Network}
\label{lcogt}

LCO is a network of queue-scheduled telescopes spread across multiple global sites that are able to provide near-continuous coverage of the night sky \citep{Brownetal2013}. Observations of TWA 3 made use of three of the southern hemisphere sites hosting 1m telescopes: Siding Springs Observatory (Australia), SAAO (South Africa), CTIO (Chile).

With the goal of detecting fast changes in the binary accretion rate, while establishing the periodicity of periastron accretion bursts, we made observations spanning $\sim$12 TWA 3A orbital periods with roughly 20 visits per orbit ($\sim$42 hr cadence). Each visit consisted of three images in each of a suite of filters. Only $U$-band photometry is discussed here; see \citet{Tofflemireetal2017b} for additional discussion of the $BVR$ photometry. All data are processed by the LCO pipeline, applying bad-pixel, bias, dark, and flat-field corrections.

Here we briefly describe our photometry and calibration schemes, referring interested readers to \citet{Tofflemireetal2017a,Tofflemireetal2017b} for a more thorough discussion. In short, after aligning and median combining the three visit images, source location and photometry are performed with SExtractor \citep{Bertin&Arnouts1996} providing time-series instrumental magnitudes. These are then input to an ensemble photometry routine following the \citet{Honeycutt1992} formalism. Output differential ensemble magnitudes are then calibrated using non-variable stars in the LCO field-of-view for which empirical or derived photometry exists. From these stars, zero-point and color transformations are derived and applied to all stars in the field. In the $U$-band we derive an uncertainty of 0.18 mag in the absolute calibration.

$U$-band photometry presented here is the combined light of the TWA 3 system (i.e. TWA 3A and TWA 3B). Flux from the two components could not be reliably separated at each epoch due to poor seeing and/or telescope focus. A subset of spatially resolved observations has shown TWA 3A to be the source of variability, and this is assumed to be the case throughout these data \citep{Tofflemireetal2017b}. Finally, mass-accretion rates are  derived from the $U$-band excess above the combined TWA 3A and 3B photospheric flux. These values represent the combined accretion rate onto TWA 3Aa and TWA 3Ab, assuming both components are accreting at the same rate. Mass accretion rate values are update from \citet{Tofflemireetal2017b} with the improved distance and stellar parameters listed in Section \ref{twa3a}. We note that due to uncertainty in our absolute flux calibration, the lowest photometric mass accretion rates may be consistent with the absence of accretion.

\section{Analysis}
\label{analysis}

In this study we wish to examine the equivalent width (EW) and velocity structure of outflow and accretion-tracing emission lines from the TWA 3A binary. Both of these measurements are complicated by the presence of the triple companion, TWA 3B. EW measurements of emission lines originating from TWA 3A are made with respect to the local continuum, to which the triple companion has a significant contribution. Direct measurements of emission line EWs from the spectra correspond to lower limits on the true TWA 3A line strengths. All three stars are also chromospherically active, with low-level emission in many of the emission lines that are used to trace accretion. To accurately determine the emission and velocity components that are strictly due to accretion, we require a measure of each star's photospheric and chromospheric contribution to a given line's shape. Complicating matters further, the relative contribution of TWA 3B is variable due to the specific seeing and pointing of each epoch. In the following subsections we describe how the relative contribution of each star is determined for each epoch with spectral line broadening functions and how these measurements enable the construction of accretion-free stellar templates, the measurement of veiling, and the correction of EW measurements.  

\subsection{Spectral-Line Broadening Functions}
\label{bf}

To decompose the relative contributions of each stellar component, we compute spectral-line broadening functions \citep{Rucinski1992}. Broadening functions are the result of a linear inversion of an observed spectrum with a zero-velocity template. Our implementation performs the inversion with singular-value decomposition following \citet{Rucinski1999}. The raw broadening function represents the function that would need to be convolved with the template in order to reproduce the observed spectrum. For instance, a simple radial-velocity shift with an otherwise perfectly matched template spectrum would result in a delta function at the shifted velocity. 

In practice, the raw broadening function is dominated by noise from uncertainty in the observed spectrum (and template spectrum if using an empirical template), and from template mismatch. And indeed, elements of the raw broadening function are not independent on scales smaller than the instrumental velocity resolution. Additionally, when convolved with a Gaussian corresponding to the template's line profile, the smoothed broadening function (hereafter BF) has the advantageous property of representing the average photospheric line profile of the observed spectrum. This makes the BF ideal for measuring the RV and \vsini \ of a source, and detecting velocity-resolved surface brightness features (i.e.\ Doppler tomography). In the case of a triple-lined system, which we examine here, the BF will have three peaks, one at each star's velocity (Figure \ref{fig:bf}). While similar to a cross-correlation function, the stable BF baseline allows for a robust measure of the relative contribution (flux ratio) of each component from the integral of each BF component. (The BF code from this study is publicly available; \citealt{saphires}\footnote{\url{https://github.com/tofflemire/saphires}}.)

One limitation of the BF in this application is that a single template is used to represent multiple stars that may have different spectral features. While other spectral decomposition methods exist that allow for component-specific templates (TODCOR, \citealt{Zucker&Mazeh1994}; TRICOR, \citealt{Zuckeretal1995}), BFs provide the most straightforward and flexible means to determine component flux ratios. 

In Appendix \ref{ap:bfs}, we explore the recovery of input flux ratios from synthetic spectroscopic binaries for different templates of varying temperature (this issue is also discussed in \citealt{Luetal2001}). In short, we find that the component flux ratio is recovered most readily by the template that provides the best radial-velocity precision for the primary star, determined from order-to-order measurement stability. Fortunately, this issue is somewhat mitigated by the small range in spectral types of the TWA 3 components (M3.0--M4.5; \citealt{Kelloggetal2017}).

With RV precision as our metric for template selection, we compute BFs using the library of empirical, high-spectral-resolution spectra of non-accreting, weak-lined T Tauri stars from \citet{Manaraetal2013,Manaraetal2017} (X-SHOOTER VIS arm; 5000--10500\AA, $R\sim$20,000)\footnote{\url{http://cdsarc.u-strasbg.fr/viz-bin/qcat?J/A+A/605/A86}}$^,$\footnote{\url{http://cdsarc.u-strasbg.fr/viz-bin/cat/J/A+A/551/A107}}. Empirical templates provide a better match to the observed spectra and produce BFs with less noise when compared to synthetic templates. We find that the TWA 9B (M3.5; \citealt{Herczeg&Hillenbrand2014}) spectrum provides the best radial-velocity precision and is selected as our template for spectral decomposition. The RV of the TWA 9B template is 10.9$\pm$0.9km s$^{-1}$, determined from a BF analysis using a 3600 K, PHOENIX template spectra \citep{Husseretal2013}. This measured RV is used to shift the empirical template to zero velocity. 

With a template selected, we compute the BF for 22 continuum normalized wavelength regions of the SALT-HRS red arm. Regions are selected from our stitched spectrum that are roughly 100\AA \ in width, masking wavelengths with emission lines and strong telluric absorption. For each region, three Gaussian profiles are then fit to the BF. After applying a 3-sigma clipping algorithm, the mean and standard deviation of these fits across spectral orders are used to measure the radial velocities and relative flux contributions of each component when well separated. For our purposes, ``well separated'' corresponds to cleanly separated BF peaks, which in practice is a separation of approximately the half-width at half maximum of the BF peak, or $\sim$10 km s$^{-1}$. While a more sophisticated profile fitting schemes may enable the decomposition of more epochs, we are ultimately limited by pointing and seeing-dependent contributions of the triple companion, for which there is no constraining information (e.g., guider camera images). Given the orbital phases of our observations, though, many of our epochs are easily decomposed (5 for SALT--HRS and 4 for FEROS), providing a representative sample of tertiary flux ratios. The average relative contribution of TWA 3B is used when correct observed EW measurements below (Section \ref{ew_meas_corr}).

\subsection{Epoch-Specific Template Construction}
\label{temp_const}

For each observation, we use the results of the BF analysis to construct an accretion-free template of the system that accounts for the variable contributions of the triple companion. Each template is constructed by adding RV-shifted versions of the original TWA 9B template (informed by the RV of BF fit peaks), which are scaled by each component's flux ratio (informed by the ratio of BF fit integrals). By scaling the three template contributions by the relative flux ratio of the TWA 3 components, the constructed template is unaffected by accretion veiling that may be present in the observed spectrum. The presence of veiling would act to dilute all three components equally, preserving their relative contribution. As the template only contain photospheric and chromospheric emission, it can be used to isolate the emission and velocity structure of accretion-tracing lines, and measure veiling.

This template construction method is flexible and computationally simple, making it ideal for this application where a unique template is required for each observation. The price of this functional ease is that two assumptions are made about the system. First, that all three stars can be adequately described by the TWA 9B template, and second, that all three stars have chromospheric emission that is constant and well described by the TWA 9B template. These assumptions are likely not problematic for TWA 3. As noted above, the three components span a small range in spectral type, and in the absence of a stellar flare, which would produce additional spectral signatures, chromospheric emission is expected to be relatively constant, to within $\sim$20\% in EW \citep{FloresSoriano2015}. 

To assess the match in chromospheric emission between the model and spectra quantitatively, we measure the EW of \hei \ 5876\AA \ in the TWA 9B template spectrum and in a spatially resolved, lower-resolution, X-SHOOTER spectrum of TWA 3B ($R\sim$11,000, \citealt{Venutietal2019}). For reference, \hei \ 5876\AA \ is the only emission line where these templates are used to isolate accretion emission in our analysis. We find an EW of -0.20$\pm$0.05\AA \ in the TWA 9B template, which is in excellent agreement with the TWA 3B measurement, -0.22$\pm$0.02\AA \ (Section \ref{ew_meas_corr}).

\subsection{Measurement of Accretion Veiling}
\label{meas_veil}

Accreting stars are known to exhibit veiling -- an additive continuum contribution that lowers the depth of photospheric absorption lines relative to the normalized continuum. In order to recover the intrinsic EW variability of accretion-tracing emission lines, the level of veiling must be determined for each epoch. The veiling continuum has been modeled as the combination of hydrogen recombination (Balmer and Paschen continuum), H$^-$ continuum, and thermal emission, originating at the base of accretion columns \citep[e.g.,][]{Calvet&Hartmann1992,Calvet&Gullbring1998,Fischeretal2011}. While the detailed spectral shape of the veiling continuum is likely to depend on the properties of the accretion flow, within the spectral rage considered here, 5500--8700\AA, \citet{Herczeg&Hillenbrand2014} find the veiling spectrum is consistent with a flat spectrum, which we adopt. 

For each epoch, we measure veiling, $V_{\rm acc} = f_{\rm acc}/f_{\rm phot}$, by performing a least squares fit of each observation with its epoch-specific, accretion-free template. In this case, $f_{\rm phot}$ corresponds to the combined continuum flux of all three stellar components. SALT and FEROS spectra are convolved with a Gaussian to lower their spectral resolution to match that of the TWA 9B template. Measurements are made in 7 regions of the spectrum from 6000 to 8700\AA \ that contain deep absorption features and smooth, well defined continua. The mean and standard error from these seven regions are presented in Tables \ref{tab:obs} and \ref{tab:feros}.

\subsection{Equivalent Width Measurement and Correction}
\label{ew_meas_corr}
For emission lines of interest we measure EWs by numerically integrating $(1-F_\lambda/F_0)$ for line profiles above a locally determined continuum flux level. Due to the limited track lengths and integration times of SALT observations, our HRS spectra do not consistently obtain detectable photospheric continuum levels at the shortest wavelengths of the blue arm. For this reason, we do not compute EWs for lines blueward of H$\beta$. For FEROS spectra we only measure H$\alpha$ and \hei \ 5876\AA \ lines for the same reason. 

In the case of \oi \ 6300\AA, we remove the terrestrial emission component by fitting a Gaussian and linear slope model to a narrow wavelength region centered at zero heliocentric
velocity. In practice, for nights with high sky emission, a two component Gaussian is required to account for the combination of adjacent spectral orders with slightly different resolution elements (see Section \ref{redarm} and Appendix \ref{ap:red}).

EW uncertainties are determined with a Monte Carlo (MC) approach. Gaussian noise at the level of adjacent continuum regions is added to the spectrum and the EW is recomputed for 10$^3$ iterations. Measurement errors are determined from the standard deviation of these simulations. EW uncertainties may be underestimated for spectral lines in the red arm ($\lambda>5500$\AA) due to a pixel-to-pixel covariance that is introduced in the image transformations described in Section \ref{redarm} and Appendix \ref{ap:red}. However, the derived uncertainties do not differ greatly between red- and blue-arm measurements (e.g., H$\alpha$ and H$\beta$). It may be the case that low-level spectroscopic features in the adjacent ``pseudo-continuum'' regions are a non-negligible source of the deviation we use in our MC procedure.

The EW measurements described above represent the TWA 3 ``combined'' EW, with contributions from three stellar components and the effect of veiling. Correcting for veiling is straight forward, but isolating the contribution from TWA 3A requires knowledge of TWA 3B's relative continuum contribution and intrinsic line shape in the wavelength regions over which the combined EW is computed. Section \ref{bf} describes how BFs are used to determine the relative flux contribution of each component when their radial velocities are well separated. We are not able to make this measurement at every epoch, so we adopt the average contribution of TWA 3B, and its standard deviation, as a representative value and uncertainty (measured from 5 SALT-HRS epochs and 4 FEROS epochs). For emission lines that originate in the accreting binary, the tertiary's continuum flux acts as an additional source of veiling, which we define as $V_{\rm 3B} = f_{\rm 3B}/(f_{\rm 3Aa} + f_{\rm 3Ab})$. The combination of the triple and accretion veiling produces an effective veiling, $V_{\rm eff} = V_{\rm acc} + V_{\rm 3B}(1 + V_{\rm acc})$.

The detailed EW correction for a given line depends on TWA 3B's spectral features. In the case of \oi \ 6300\AA, where the triple's spectrum is relatively featureless, we treat the triple as a flat veiling continuum ($V_{\rm 3B}$) and correct for it with the following equation:
\begin{equation}
    EW_{\rm 3A_{[OI]}} = EW_{\rm Obs_{[OI]}}(1+V_{\rm eff}).
\label{eqn:ew_oi}
\end{equation}

For accretion-tracing lines (\hal, H$\beta$, \hei \ 5876\AA), each star has a low-level chromospheric emission component. The EW measurements of accretion-tracing emission lines in single stars also contain this chromospheric component, so to make a direct comparison with previous studies, we correct the combined EW measurement for the triple's veiling continuum and its chromospheric contribution (leaving only the binary's accretion and chromospheric emission). We apply the same veiling continuum correction but then apply an additional term to remove the triple's chromospheric EW contribution. The explicit equation is as follows:
\begin{equation}
    EW_{\rm 3A_{i}} = EW_{\rm Obs_{i}}(1+V_{\rm eff}) - EW_{\rm 3B_{Chro,i}} V_{\rm 3B},
\label{eqn:ew}
\end{equation}
where $i$ refers to different accretion tracing lines (\hal, H$\beta$, \hei \ 5876\AA), $EW_{\rm 3B_{Chro,i}}$ is the EW of TWA 3B's chromospheric emission for the corresponding accretion-tracing emission line. $EW_{\rm 3B_{Chro,i}}$ is determined from a spatially resolved spectrum of TWA 3B with the X-SHOOTER spectrograph on the VLT, provided as a reduced data product from the ESO Archive\footnote{\url{http://www.eso.org/rm/api/v1/public/releaseDescriptions/70}}, and also presented in \citet{Venutietal2019}. We measure chromospheric EWs of  -4.73$\pm$0.09\AA, -4.80$\pm$0.04\AA, and -0.22$\pm$0.02\AA, for \hal, H$\beta$, and \hei \ 5876\AA, respectively. (The spectral resolution of this observation, $R\sim$ 11,000,  is too low to be used as our BF template.) Here we have assumed that the chromospheric contribution from TWA 3B is constant, which is generally true within $\sim$$<$20\% for magnetically active stars \citep{FloresSoriano2015}, and much smaller than the accretion variability observed (see Section \ref{ews}). The corrected EW measurements are presented in Tables \ref{tab:obs} and \ref{tab:feros}. 

\begin{figure}[!t]
  \centering
  \includegraphics[keepaspectratio=true,scale=1.1]{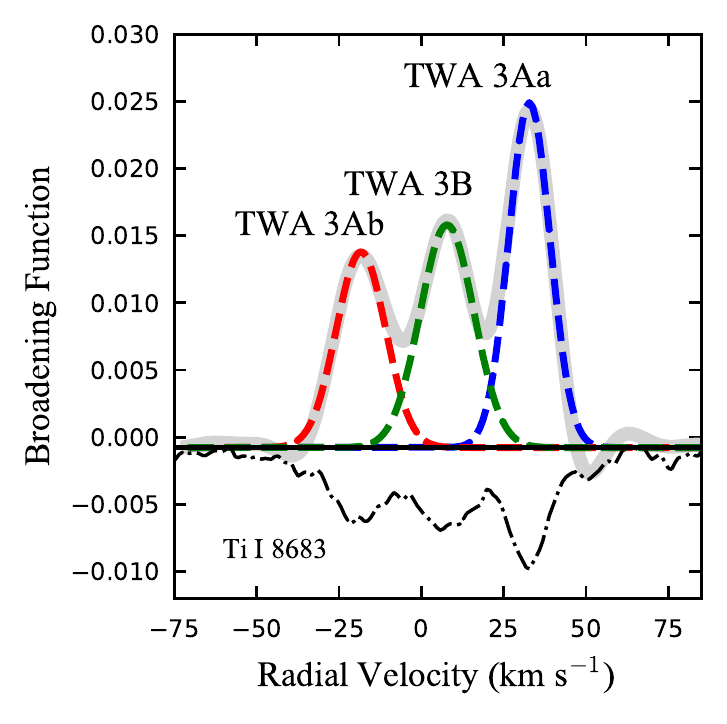}
  \caption{Broadening function (BF) for a TWA 3 SALT-HRS observation where all three components can be easily decomposed. The BF is shown in grey with a Gaussian fit to each component (labeled). At the bottom of the figure, the dot-dashed line presents the scaled line profile of Ti {\scshape i} 8683, highlighting the BF's reconstruction of the average photospheric absorption line profile.}
  \label{fig:bf}
\end{figure}

\begin{figure}[!t]
  \centering
  \includegraphics[keepaspectratio=true,scale=1.02]{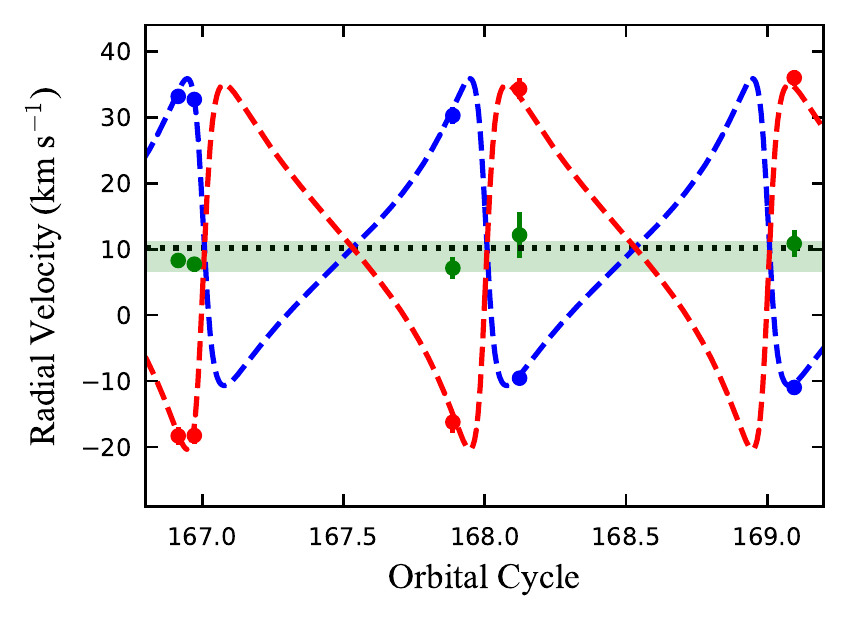}
  \caption{Radial-velocity measurements of the three TWA 3 stars from 5 SALT--HRS epochs where all three components are widely separated in velocity. Blue, red and green points mark the primary (TWA 3Aa), secondary (TWA 3Ab), and tertiary (TWA 3B), respectively. The $\gamma$ velocity (dotted line), orbital solution (dashed lines), and green band, representing the standard deviation of CfA RV measurements of TWA 3B, are from \citet{Kelloggetal2017}.}
  \label{fig:rvs}
\end{figure}

\begin{figure*}[!t]
  \centering
  \includegraphics[keepaspectratio=true,scale=0.95]{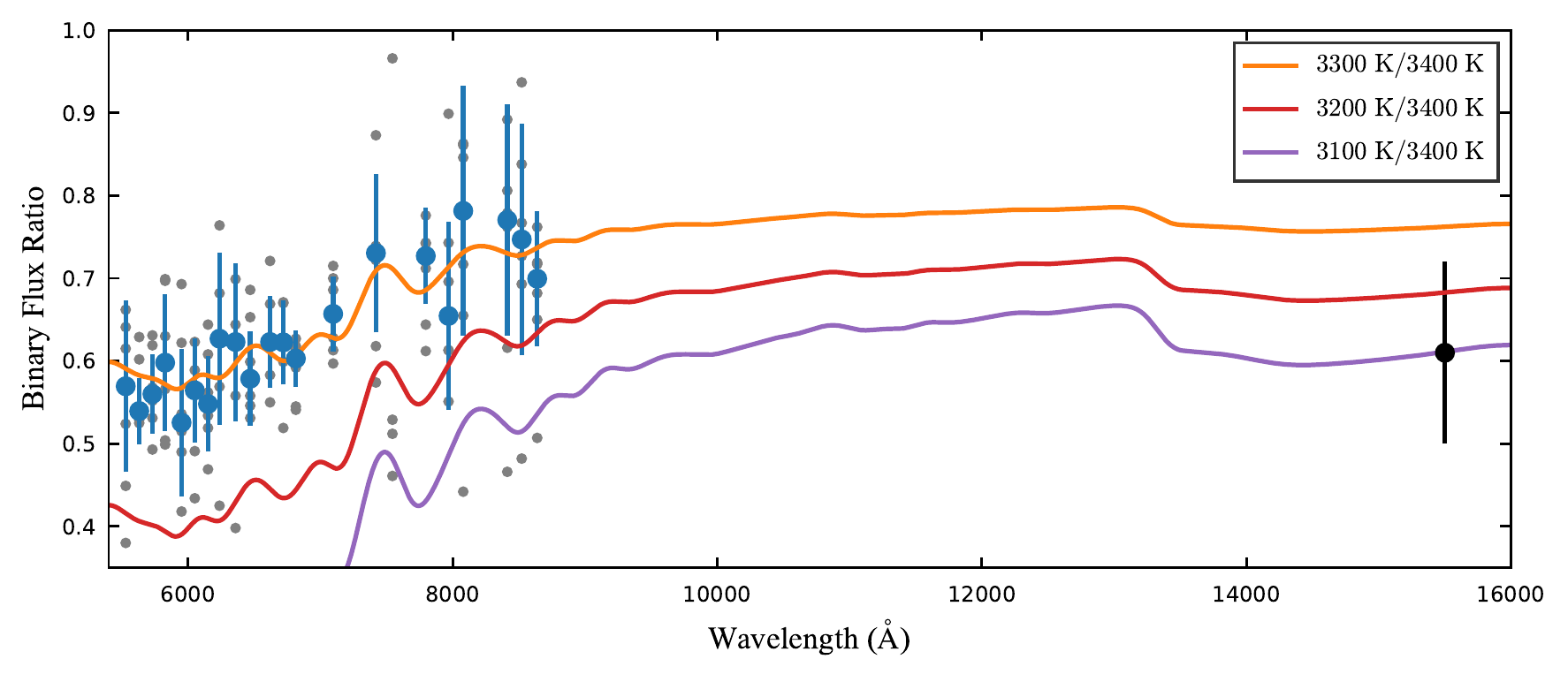}
  \caption{TWA 3A binary flux ratio ($f_{\rm 3Ab}/f_{\rm 3Aa}$) as a function of wavelength. Measurements are derived from observations where all three stellar components (TWA 3Aa, 3Ab, 3B) can be distinguished by our broadening function analysis. Grey points are measurements from individual epochs. Blue points are the median of the wavelength regions with standard deviations represented as the error bars. Solid lines overlays the flux ratio of synthetic templates \citep{Baraffeetal2015}. The black point marks the flux ratio determined by \citet{Kelloggetal2017} from H-band spectra using TODCOR.}
  \label{fig:fratio}
\end{figure*}

\begin{figure}[!t]
  \centering
  \includegraphics[keepaspectratio=true,scale=1.02]{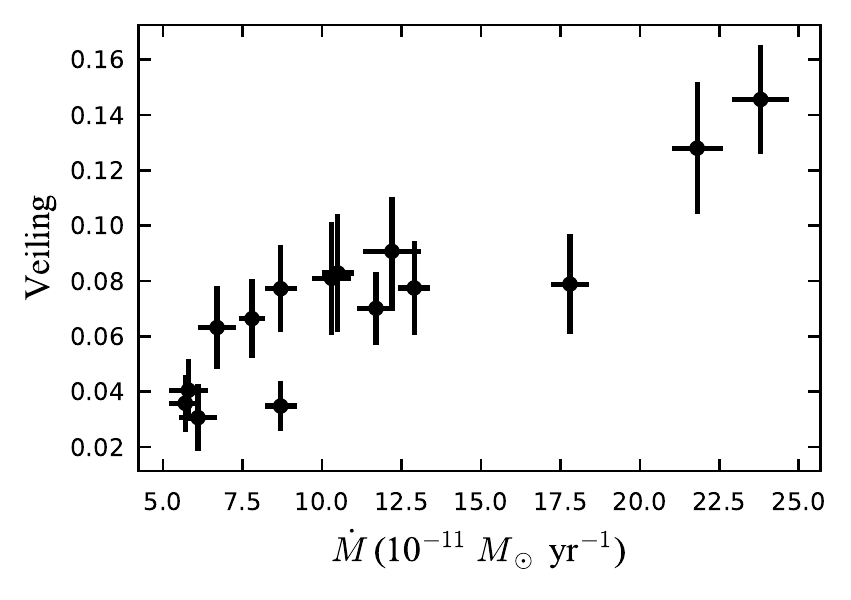}
  \caption{Veiling measurements from SALT--HRS red arm plotted against the photometrically derived mass accretion rate.}
  \label{fig:veiling}
\end{figure}

\section{Results}
\label{results}

In this section we begin by presenting the results of our spectral decomposition, including RV measurements, component flux ratios, and veiling measurements. We then present EW measurements for select lines, followed by an analysis of the velocity structures of emission lines. 

\subsection{Spectral Decomposition}
\label{specdec}

The BF analysis described in Section \ref{bf} provides the radial velocity and flux ratio of all three stellar components, when they are adequately separated in velocity. Figure \ref{fig:bf} presents the BF for one epoch with well-separated components. The velocity structure of a deep, isolated photospheric absorption line, Ti {\scshape i} 8683, is also included to demonstrate the BFs representation as the average absorption line profile. The RV measurements of 5 SALT-HRS epochs with large velocity separations are presented in Figure \ref{fig:rvs}. These measurements are in excellent agreement with the over plotted orbital solution from \citet{Kelloggetal2017}. 

Figure \ref{fig:fratio} displays the TWA 3A binary flux ratio for well-separated observations in grey points as a function of the region's central wavelength. Blue points and their error bars represent the median and standard deviations of individual measurements. A wavelength dependence of the TWA 3A binary flux ratio can be seen, which is consistent with the secondary having a lower temperature. The solid  lines overlay the flux ratio of synthetic templates \citep[log($g$) $=$ 4.5;][]{Baraffeetal2015}, scaled by the radius of each component. At H-band wavelengths, this model predicts a larger flux ratio than what was measured in by \citet{Kelloggetal2017} from H-band Keck-NIRSPEC spectra using TODCOR (black circle). The discrepancy between these measurements is likely due to systematic offsets between methods and the templates used. This discrepancy aside, both results support a small difference in the temperature of the binary components ($\lesssim$200 K).

Turning to the triple-to-binary flux ratio, we find the contribution from TWA 3B is significant (Figure \ref{fig:bf}) and variable. On average it contributes 33\% of the SALT-HRS red arm flux with a standard deviation of 8\%, corresponding to $V_{3B}$=0.52$\pm$0.19. It is only for these 5 measurements that we are able to make robust measurements of all three components but we assume they are representative of all 15 spectra. Following the same procedure from 4 of the 9 FEROS observations, the TWA 3B contribution is 0.51$\pm$0.11 ($V_{3B}$=1.12$\pm$0.43).

Constructing an accretion-free template for each epoch (see Section \ref{temp_const}), we test for the presence of spectral line veiling (method described in Section \ref{meas_veil}). We detect veiling at each epoch and find it is well-correlated with the photometrically determined mass accretion rate, as presented in Figure \ref{fig:veiling}.

\begin{figure}[!t]
  \centering
  \includegraphics[keepaspectratio=true,scale=1.1]{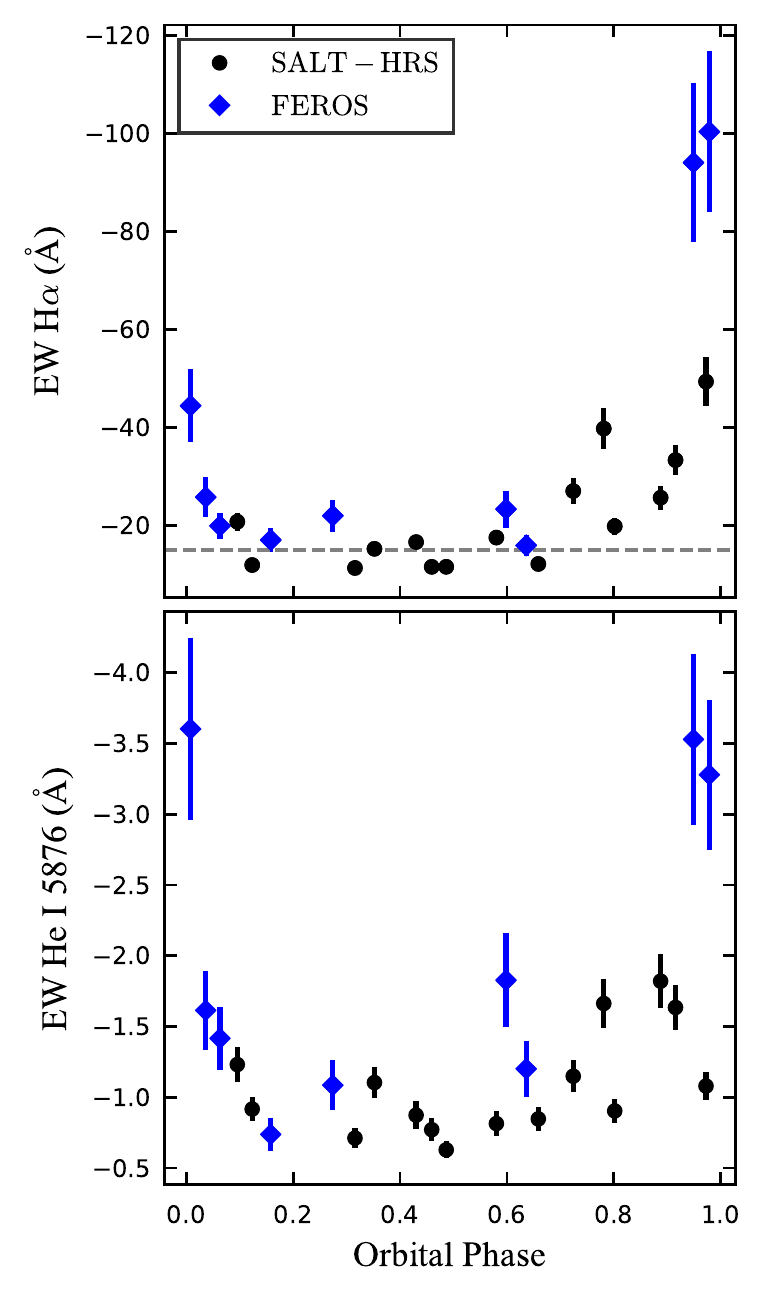}
  \caption{Equivalent widths (EWs) of \hal \ and \hei \ 5876\AA \ as a function of orbital phase. Black circles are SALT--HRS observations; blue diamonds are FEROS observations. The large uncertainty on high EW measurements is due to the uncertainties on the contribution from TWA 3B and do not reflect the uncertainties of the raw EW measurements. The horizontal gray line marks the \hal \ EW boundary between chromospheric- and accretion-based emission determined by \citet{Fangetal2009}.}
  \label{fig:ew_phase}
\end{figure}

\begin{figure}[!h]
  \centering
  \includegraphics[keepaspectratio=true,scale=1.05]{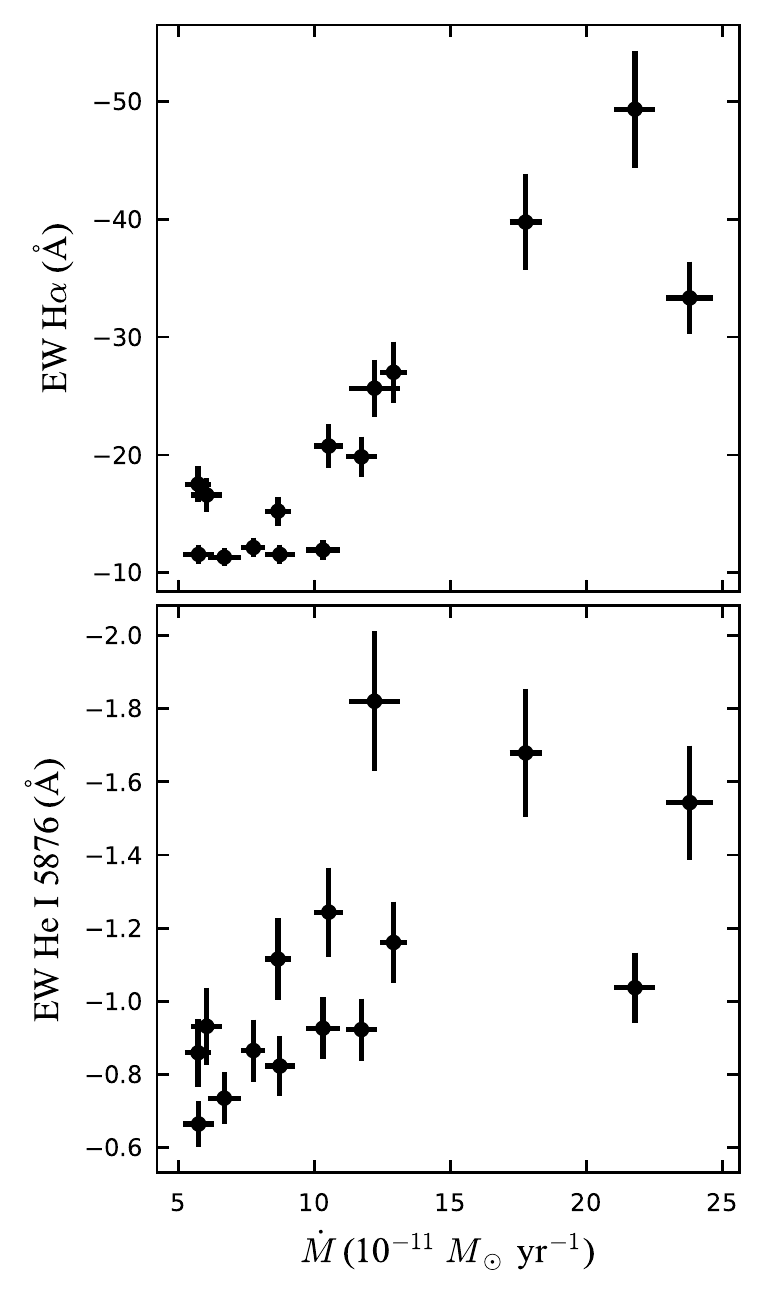}
  \caption{Equivalent widths (EWs) of \hal \ and \hei \ 5876\AA \ from SALT--HRS observations compared to a photometrically determined mass-accretion rates. Mass-accretion rate for each equivalent-width measurement is the linear interpolation of the adjacent photometric observations.}
  \label{fig:ew_md}
\end{figure}

\subsection{Equivalent-Width Variability}
\label{ews}

Figure \ref{fig:ew_phase} presents the EW measurements for the accretion-tracing emission lines H$\alpha$ and \hei \ 5876\AA \ as a function of orbital phase. Measurements have been corrected for the effects of veiling and the contribution of TWA 3B (see Section \ref{ew_meas_corr}). Black and blue points mark SALT--HRS and FEROS observations, respectively. Note that the large error bars on high EW FEROS measurements result from a large uncertainty on the contribution of TWA 3B that scales with the measured EW; these measurements do not have a higher relative uncertainty than the lower EW measurements.

Seen most prominently in H$\alpha$, the increase in EW near periastron confirms the presence of enhanced accretion events previously observed with photometry in \citet{Tofflemireetal2017b}. This behavior is also present in Figure \ref{fig:ew_md} where the \hal \ and \hei \ 5876\AA \ EWs are plotted against the photometrically derived mass-accretion rate. (Mass-accretion rate is interpolated from adjacent LCO observations.) Some degree of scatter in this figure is due to the non-simultaneity of the spectral and photometric measurements.  

During quiescence, the H$\alpha$ EW hovers between $\sim$ $-$10 and $-$20\AA. For M3-4 spectral types (similar to TWA 3A), \citet{Fangetal2009} define an H$\alpha$ EW of $-$15\AA \  as the boundary between classical and weak-lined T Tauri stars (WTTSs), meant to separate accreting and non-accreting systems. This supports the very low veiling and accretion rates measured outside of periastron. While a similar CTTS--WTTS boundary does not exist explicitly in the literature for \hei \ 5876\AA \ EWs, \citet{Manaraetal2013} find an average EW of -0.6\AA\ for WTTS with M3 and M4 spectral types, signaling a consistent behavior for both \hei\ and \hal. For reference, the EW of \hal \ and \hei \ in the TWA 9B template fall below both of these CTTS--WTTS boundaries.

Though not presented graphically, we also measure the EWs of \oi \ 6300\AA. \oi \ is typically assumed to probe diffuse outflowing material in disk or accretion driven winds \citep{Hartiganetal1995}. After correcting for veiling the TWA 3B continuum, we find the EW is consistent with being constant across our measurements ($\chi^2_\nu$=0.8) with an average EW of -0.48$\pm$0.10\AA. \oi \ is discussed further in Section \ref{oi} below where its velocity profiles are presented. 

\begin{figure*}[!h]
  \centering
  \includegraphics[keepaspectratio=true,scale=1.2]{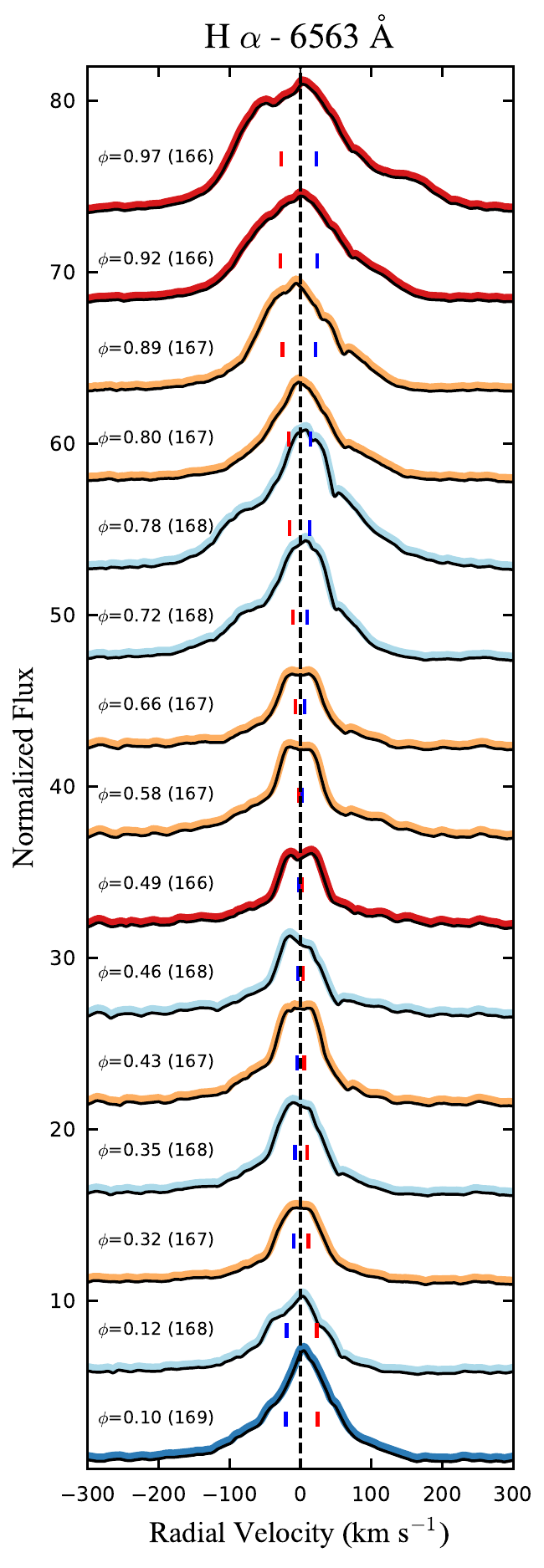}
  \includegraphics[keepaspectratio=true,scale=1.2]{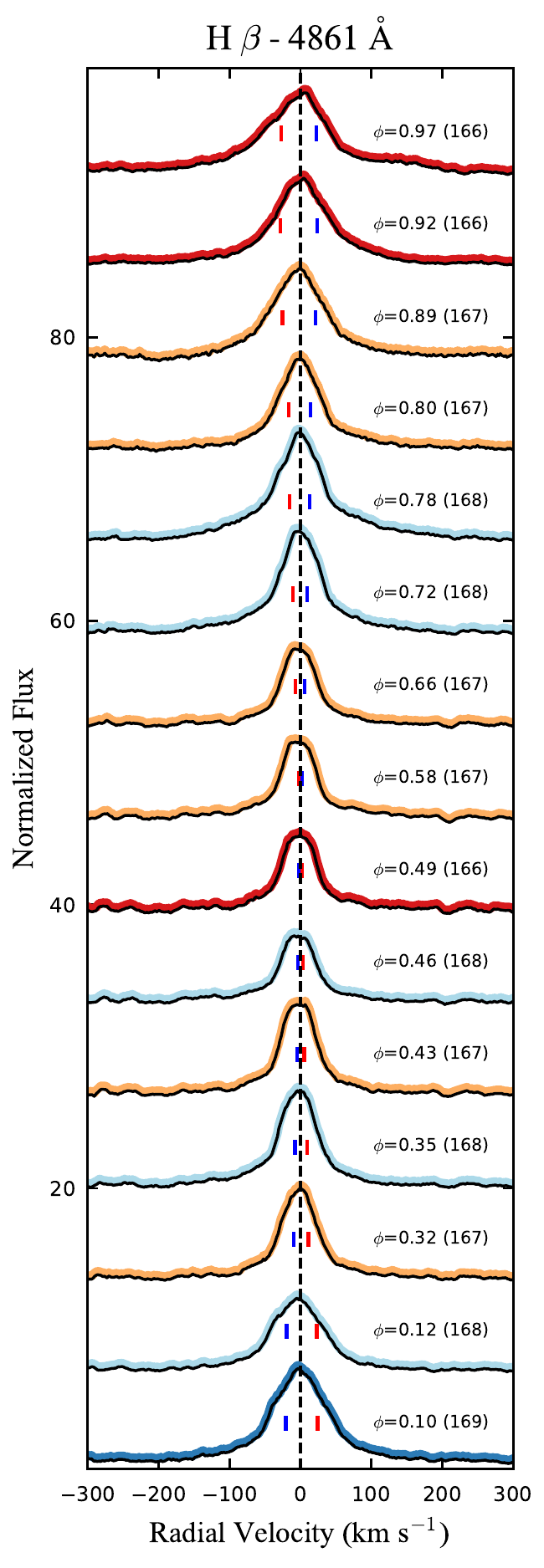}
  \caption{Velocity structure of continuum-normalized Balmer emission lines H$\alpha$ and H$\beta$ from SALT--HRS observations. Velocities are presented with respect to the TWA 3A center-of-mass velocity (also marked with a dashed line). Spectra are ordered from bottom to top by increasing orbital phase, which is labeled adjacent to each spectrum. The line color marks the orbital cycle of the observation, which is listed in the label parenthetical. Vertical dashes associated with each spectrum mark the velocities of the primary and secondary stars in blue and red, respectively.}
  \label{fig:balmer}
\end{figure*}
\renewcommand{\thefigure}{\arabic{figure} (cont.)}
\addtocounter{figure}{-1}
\begin{figure*}[!hp]
  \centering
  \includegraphics[keepaspectratio=true,scale=1.2]{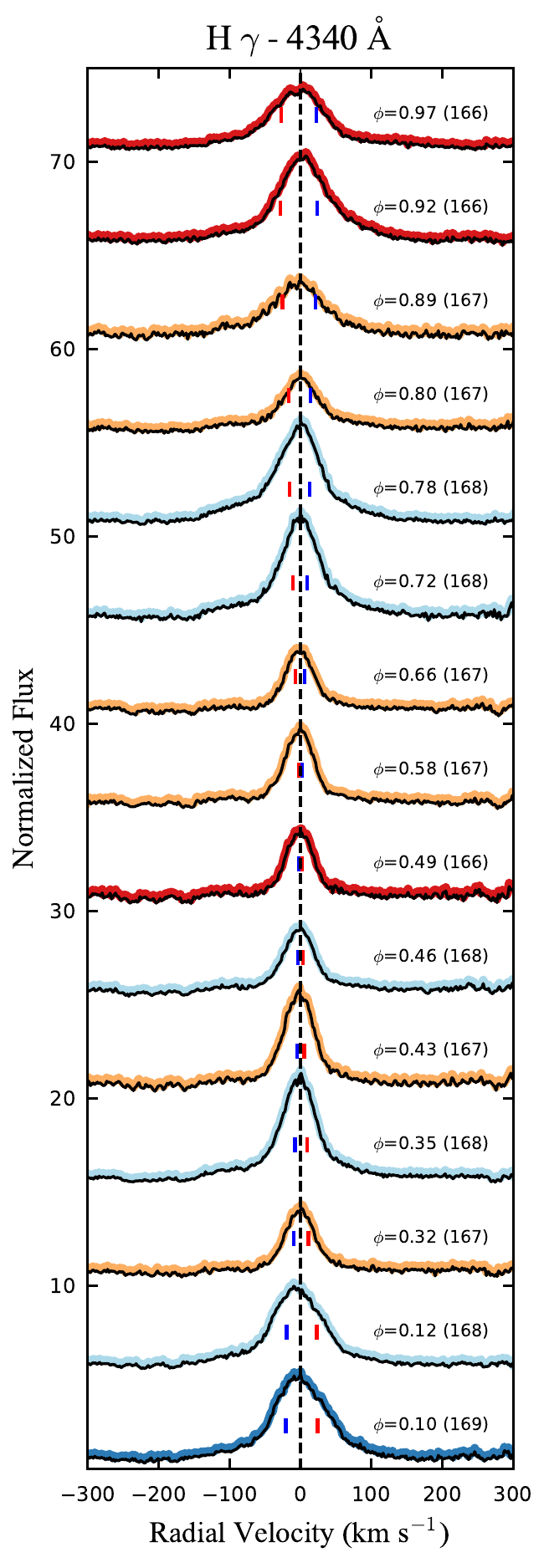}
  \includegraphics[keepaspectratio=true,scale=1.2]{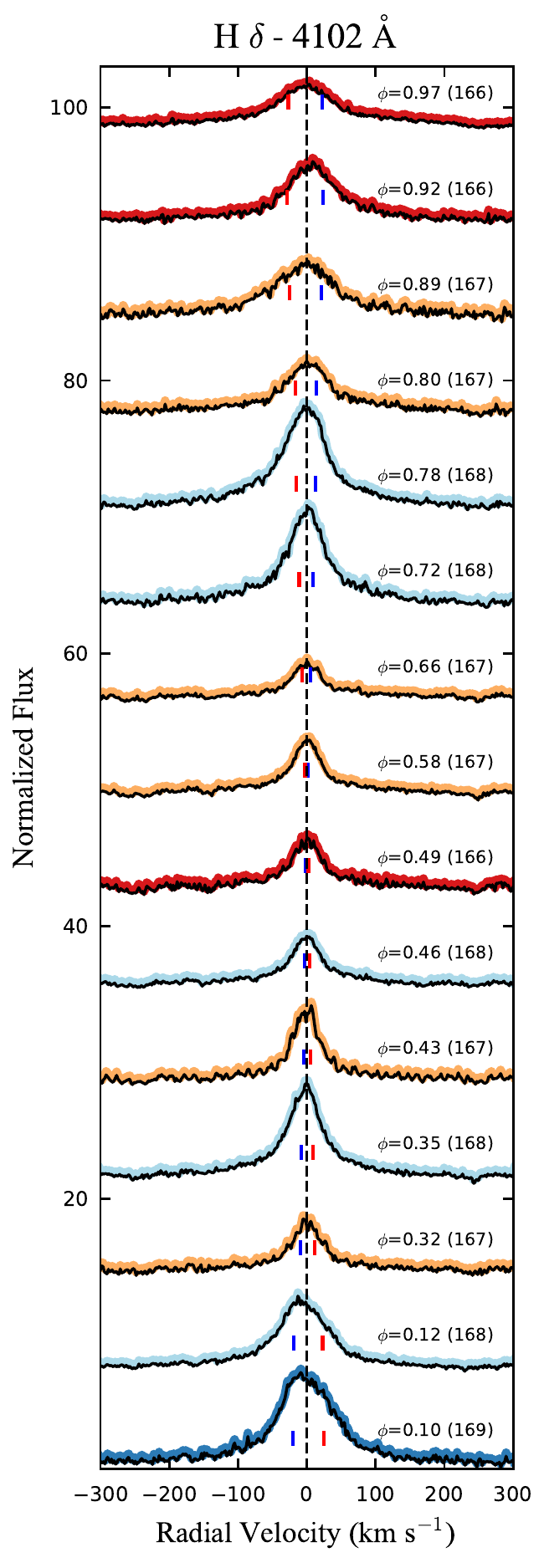}
  \caption{Velocity structure of normalized Balmer emission lines H$\gamma$ and H$\delta$ from SALT--HRS observations. The stellar continuum is not consistently detected at these wavelengths; normalization is with respect to the sky/detection background. Only lines shapes are comparable between these spectra. (See top panels for a full description of figure details.)}
\end{figure*}
\renewcommand{\thefigure}{\arabic{figure} (cont.)}
\addtocounter{figure}{-1}

\begin{figure}[!hp]
  \centering
  \includegraphics[keepaspectratio=true,scale=1.2]{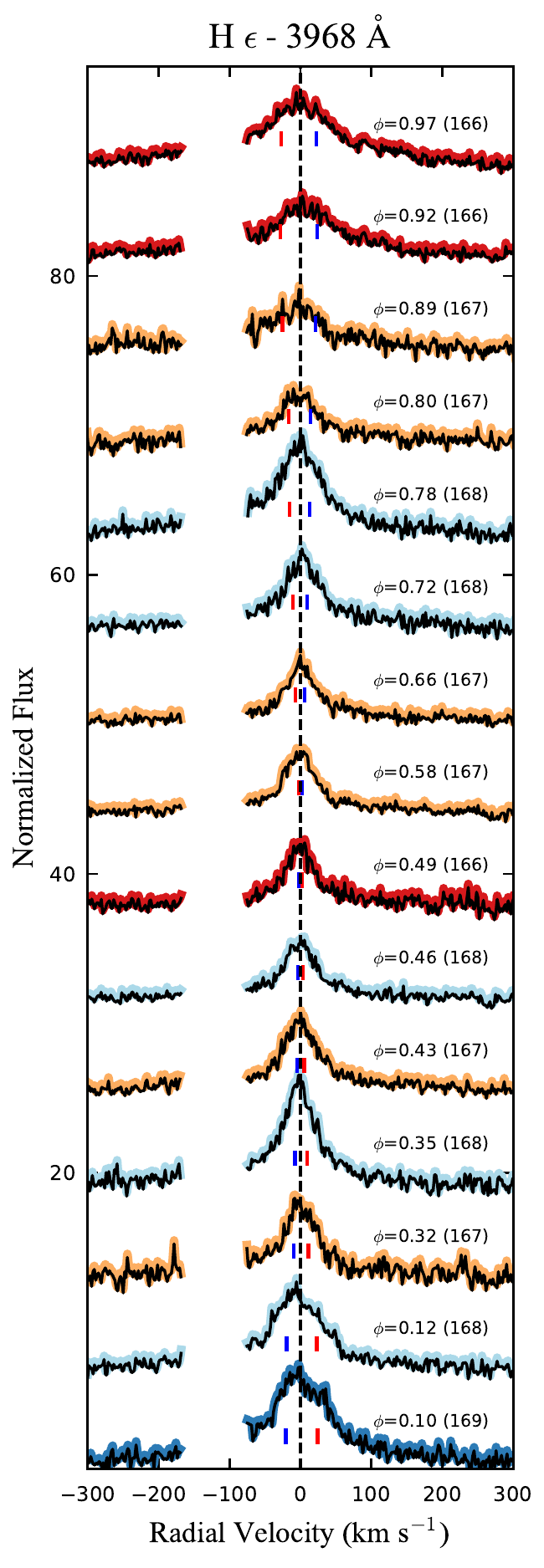}
  \caption{Velocity structure of Balmer emission line H$\epsilon$ from SALT--HRS observations. Gap at negative velocities excludes \caii \ H. Stellar continuum is not consistently detected at these wavelengths; normalization is with respect to the sky/detection background. Only lines shapes are comparable between these spectra. (See top panels for a full description of figure details.)}
\end{figure}
\renewcommand{\thefigure}{\arabic{figure}}

\begin{figure}[!hp]
  \centering
  \includegraphics[keepaspectratio=true,scale=1.2]{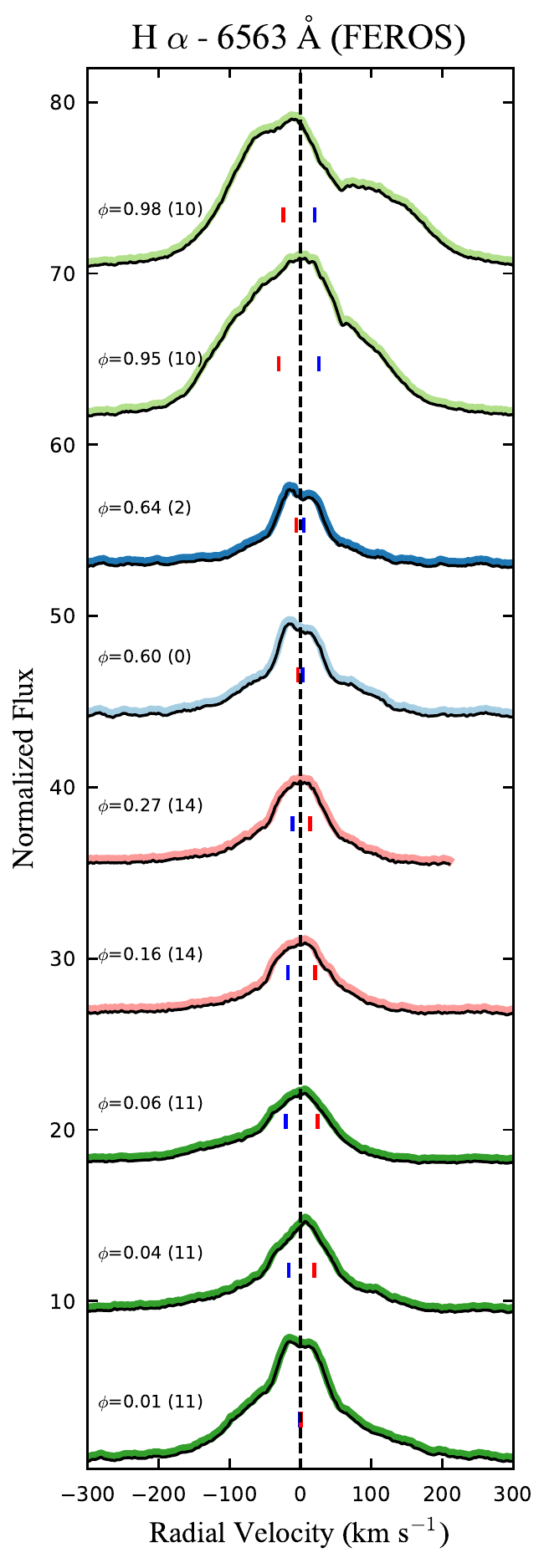}
  \caption{Velocity structure of \hal \ from archival FEROS observations. (Figure has the same layout as Figure \ref{fig:balmer}, see a full description there.)}
\label{fig:haf}
\end{figure}
\renewcommand{\thefigure}{\arabic{figure}}

\subsection{Spectral-Line Velocity Structures}
\label{velocity}

To examine the kinematic structure of accretion- and outflow-tracing spectral lines, the following subsections present the radial-velocity profiles of select spectral features in the TWA 3A binary center-of-mass rest frame. Figures \ref{fig:balmer}, \ref{fig:haf}, \ref{fig:hei}, and \ref{fig:heiferos} present continuum normalized spectral lines in the following format: spectra are ordered from bottom to top by increasing orbital phase, which is presented to the left or right of each curve. The line color and parenthetical in each label displays the orbital cycle in which the observation was made. For each spectrum the primary and secondary stellar radial velocity is presented as a vertical blue and red line, respectively. Continuum normalization has been performed locally for each line. For brevity, we only discuss H Balmer lines, \hei \ 5876\AA, and \oi \ 6300\AA \ in the following subsections but include figures and brief descriptions of \caii \ H and K and \hei \ 4471\AA \ in Appendix \ref{ap:lines}.

\subsubsection{H Balmer Series}
\label{bamler}

Figure \ref{fig:balmer} presents the velocity structure of H$\alpha$, H$\beta$, H$\gamma$, H$\delta$, and H$\epsilon$. Inspection of the H$\alpha$ and H$\beta$ panels reveals strong increases in their emission strengths near periastron, mirroring the EW behavior in Figure \ref{fig:ew_phase} above, as expected. The velocity widths and complexity of line structure are also seen to increase with accretion rate, similar to that seen in DQ Tau accretion bursts \citep{Basrietal1997}. Near apastron (spectra in middle of figure) emission is relatively narrow and stable, and is likely to be dominated by chromospheric emission (e.g., \citealt{Houdebineetal2012}). The consistent ``notch'' feature in the red wing of \hal \ near $\sim$50 \kms \ is a terrestrial absorption feature.

The following panels present H$\delta$, H$\gamma$, and H$\epsilon$. Here we again find narrow emission near apastron that becomes broader near periastron (top and bottom). Emission is mostly centered and symmetric with exceptions at $\phi$=0.10, 0.12, and 0.92 where there are asymmetries that skew toward the primary stellar velocity. We note that low levels of the stellar continua at these wavelengths are below our sky/background detection limit. Thus the spectra in these panels are normalized to the sky/background noise level. 
This makes the relative amplitude of lines dependent on the conditions of individual observations (e.g., atmospheric transparency) rather than their intrinsic strength with respect to the continuum. 

Figure \ref{fig:haf} presents the \hal \ emission from archival FEROS spectra. These spectra span $\sim$15 orbital periods that predate our SALT--HRS observations by $\sim$150 orbital cycles. These data allow us to comment on the  longevity of observed trends while increasing the phase coverage and number of periastron accretion outbursts observed. In these epochs the same trends hold, with the EWs, velocity widths, and line structure increasing near periastron passage. 

\subsubsection{He I 5876\AA}
\label{hei_decomp}

\begin{figure*}[!hp]
  \centering
  \includegraphics[keepaspectratio=true,scale=1.2]{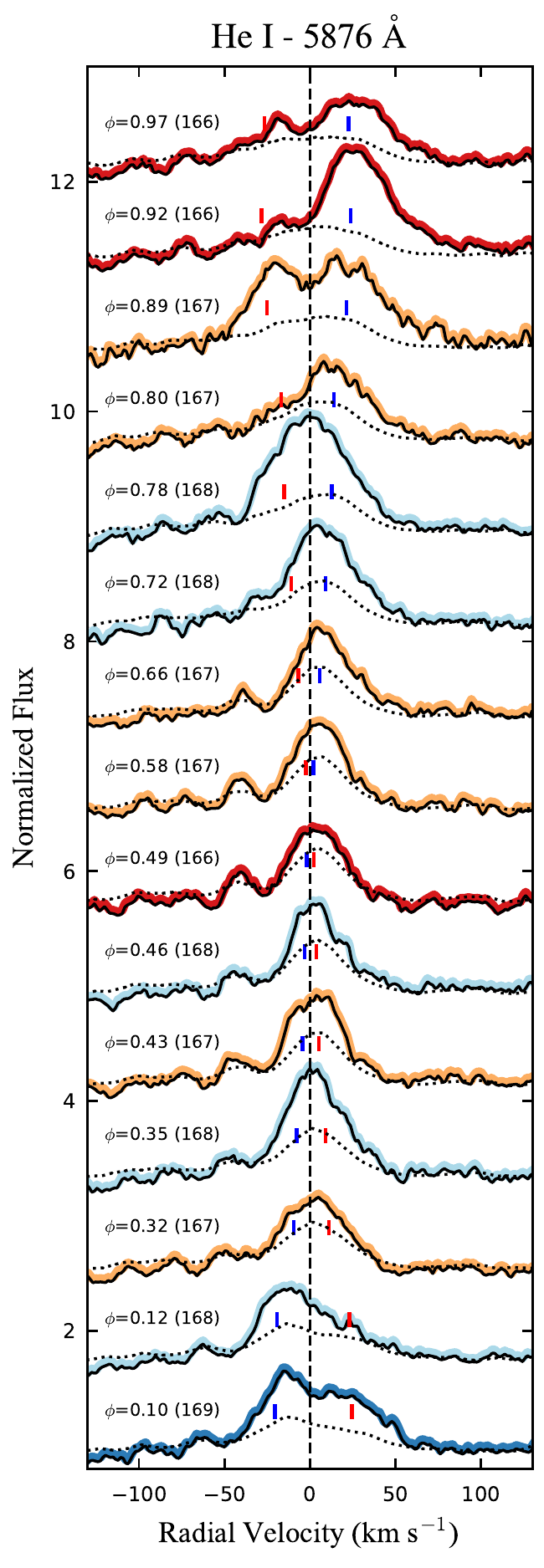}
  \includegraphics[keepaspectratio=true,scale=1.2]{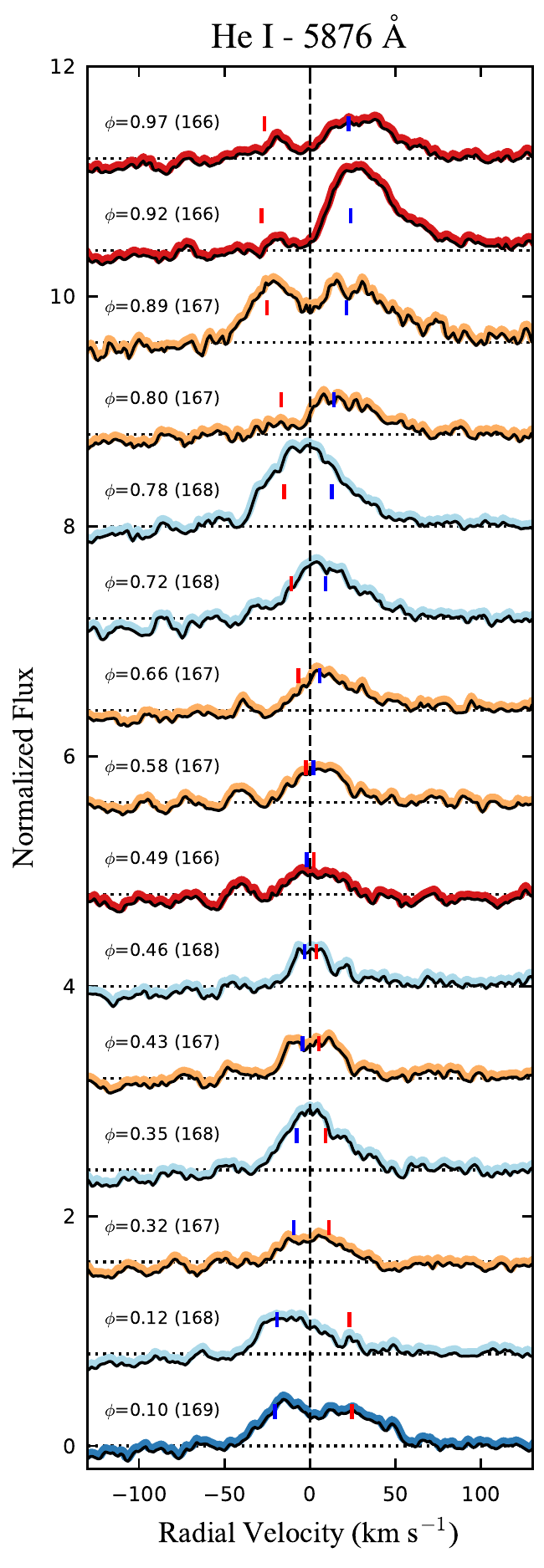}
  \caption{Velocity structure of \hei \ emission line from SALT--HRS observations. In the left panel, the observed spectrum is in the colored line and a model of the photospheric and chromospheric emission is shown with the dotted line (see Section \ref{temp_const}). The right figure subtracts the model leaving only emission from accretion. (Figure has the same layout as Figure \ref{fig:balmer}; see a full description there.)}
  \vspace{5pt}
  \label{fig:hei}
\end{figure*}

\begin{figure*}[!hp]
  \centering
  \includegraphics[keepaspectratio=true,scale=1.2]{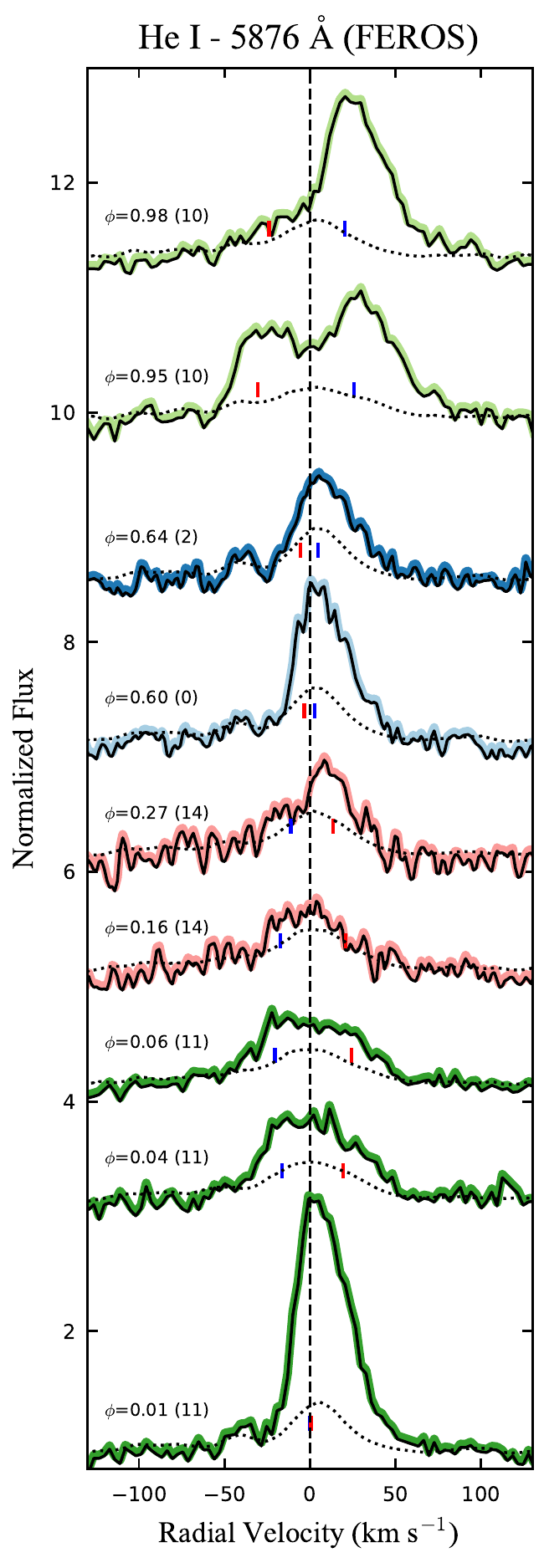}
  \includegraphics[keepaspectratio=true,scale=1.2]{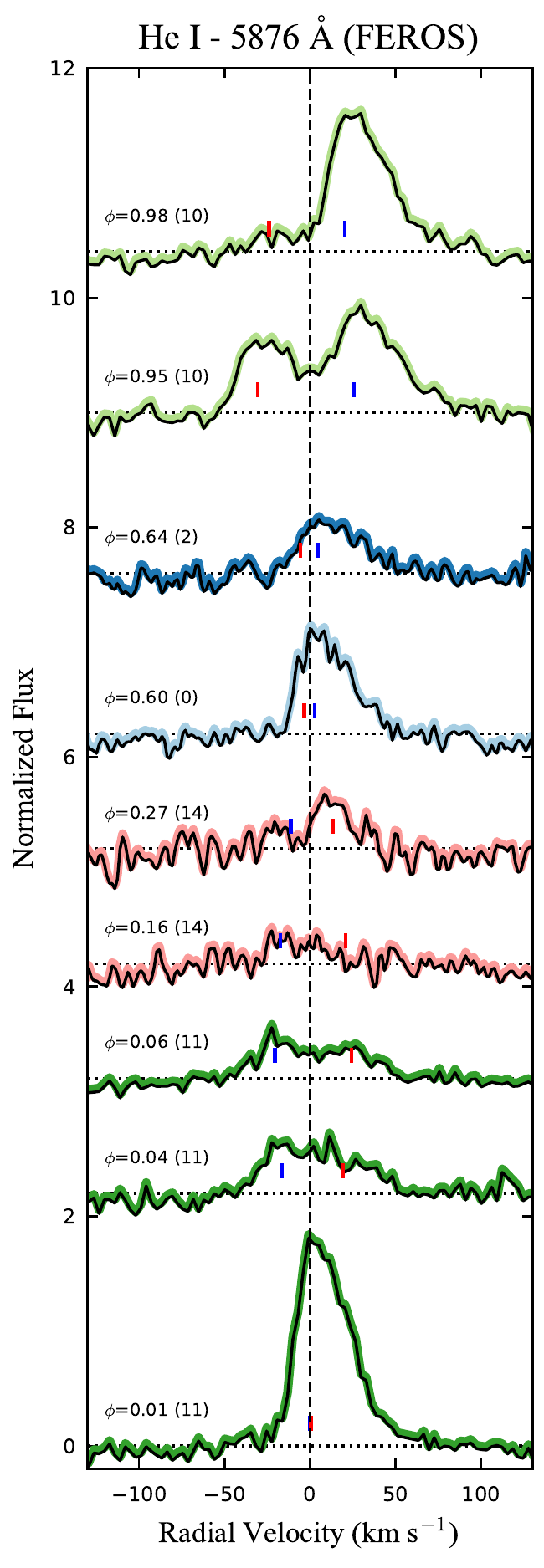}
  \caption{Velocity structure of \hei \ emission line from archival FEROS observations. In the left panel, the observed spectrum is in the colored line and a model of the photospheric and chromospheric emission is shown in with the dotted line (see Section \ref{temp_const}). The right figure subtracts the model leaving emission from accretion. (Figure has the same layout as Figure \ref{fig:balmer}; see a full description there.)}
  \label{fig:heiferos}
  \vspace{5pt}
\end{figure*}

\hei \ 5876\AA \ is an accretion-tracing emission line that is assumed to trace the base of accretion shocks (e.g., \citealt{Johns-Krull&Basri1997}). It is unique among most accretion-tracing emission lines in that it has a relatively narrow profile and suffers less from self-absorption and the optical depth effects that complicate the analysis of \hal, (for instance, \citealt{Muzerolleetal1998,Dahm2008,Alencaretal2012}). In the left panels of Figures \ref{fig:hei} and \ref{fig:heiferos} we present the observed velocity profiles of \hei \ 5876\AA \ from the SALT--HRS and FEROS spectra, respectively. In these spectra  we find similar trends to those seen in H$\alpha$ where emission strength, velocity width, and velocity structure increase near periastron passage. Many epochs also show a double peak or multi-component morphology. The comparison between SALT--HRS and FEROS observations also shows a consistent phase-dependent behavior separated by $\sim$150 orbital periods.

Strikingly, in many of the highest EW observations, the dominant component of emission is centered on the primary's stellar velocity. The observations at $\phi$=0.92 (Figure \ref{fig:hei}) and $\phi$=0.98 (Figure \ref{fig:heiferos}) are the most dramatic examples. The same qualitative behavior can be seen in the \caii\ H and K and \hei\ 4471\AA\ emission line profiles included in Appendix \ref{ap:lines} (Figure \ref{fig:cahkh}). (As discussed in Section \ref{salt} and Appendix \ref{ap:lines}, these lines are excluded from our current analysis due to low signal in the line's emission or in the surrounding continuum.) The same behavior is not seen in \hal\ profiles, which we attribute to complex radiative transfer through intervening accretion columns \citep{Muzerolleetal2001,Thanathibodeeetal2019} and/or outflows \citep[e.g.,][]{Dupreeetal2012} that shape the line's emergent velocity structure more than the RVs of the stellar components. The difference between \hal\ and \hei\ can be seen most clearly comparing the $\phi$=0.98 epoch in Figures \ref{fig:haf} and \ref{fig:heiferos}. With a low-optical depth line like \hei\ 5876\AA, we have the prospect of decomposing the flux, and hence, accretion rate onto each star. We focus the rest of this section on the determining the contribution of each star's accretion and chromospheric emission to the observed \hei\ line profile.

\hei \ emission from TWA 3A is weak, and has the potential to be contaminated by chromospheric emission. To isolate the emission from accretion processes we remove the chromospheric emission by subtracting a continuum-normalized, epoch-specific template (see Section \ref{temp_const}). Each template is veiled to match it's given epoch, and is represented by the black dotted line in the left panels of Figures \ref{fig:hei} and \ref{fig:heiferos}. The right panels of these figures present the subtracted spectrum, which only includes emission from accretion processes (given our assumptions above). 

In the right panels of Figures \ref{fig:hei} and \ref{fig:heiferos} we find that near apastron (middle spectra), much of the \hei \ emission can be represented by chromospheric emission, signifying little ongoing accretion outside of the discrete bursts. This agrees qualitatively with the narrow, double-peaked \hal \ profiles at the same orbital phases (Figure \ref{fig:balmer}). 

Near periastron, we find that \hei \ emission strength continues to favor the primary stellar velocity in these subtracted spectra. The line profile also skews toward positive radial velocities, particularly at high-emission epochs (Figure \ref{fig:hei}, $\phi$=0.92; Figure \ref{fig:heiferos}, $\phi$=0.01,0.98), which is common in single CTTS \citep[e.g.,][]{Johns-Krull&Basri1997,Alencaretal2018}.

To assign \hei \ flux, and therefore a relative accretion rate, to each star, we develop a simple model to describe the ``ownership'' of emission to each star as a function of radial velocity. We begin by fitting a skewed Gaussian profile to three chromosphere-subtracted spectra with high emission strengths that either cleanly originate from one star (HRS - $\phi$=0.92; FEROS - $\phi$=0.98) or come from an epoch where the stellar velocities are overlapping (FEROS - $\phi$=0.01). The skewed distribution has the following form:
\begin{equation}
    p(v) = Ae^{-\frac{(v-v_0)^2}{2w^2}}\left[ 1+{\rm erf} \left( \frac{\alpha(v-v_0)}{w} \right) \right],
\end{equation}
where $A$ is the profile amplitude, $v$ is the RV, $v_0$ is the profile center, $w$ is the profile width, $\alpha$ is the skewness parameter, and erf is the error function. The peak of a skewed Gaussian differs from $v_0$, so we parameterize it as $v_0 = v_\star + v_{\rm off}$, where $v_\star$ is the known stellar velocity and $v_{\rm off}$ is the offset from the distribution's nominal center. The mean and standard deviation of these best-fit parameters comprise our ``base model'' for a single star. Specifically, $w$=30$\pm$4 km s$^{-1}$, $v_{\rm off}$=$-$10$\pm$2 km s$^{-1}$, and $\alpha$=2.6$\pm$0.3. For each spectrum, this model is fit at each star's RV where the profile location, width, and skewness are only allowed to vary within 2$\sigma$ of the base model, while the amplitude is left free. (Spectra with nearly overlapping RV are excluded from our analysis, i.e., orbital phases $\phi$=0.5--0.6.)

\begin{figure}[!t]
  \centering
  \includegraphics[keepaspectratio=true,scale=1.1]{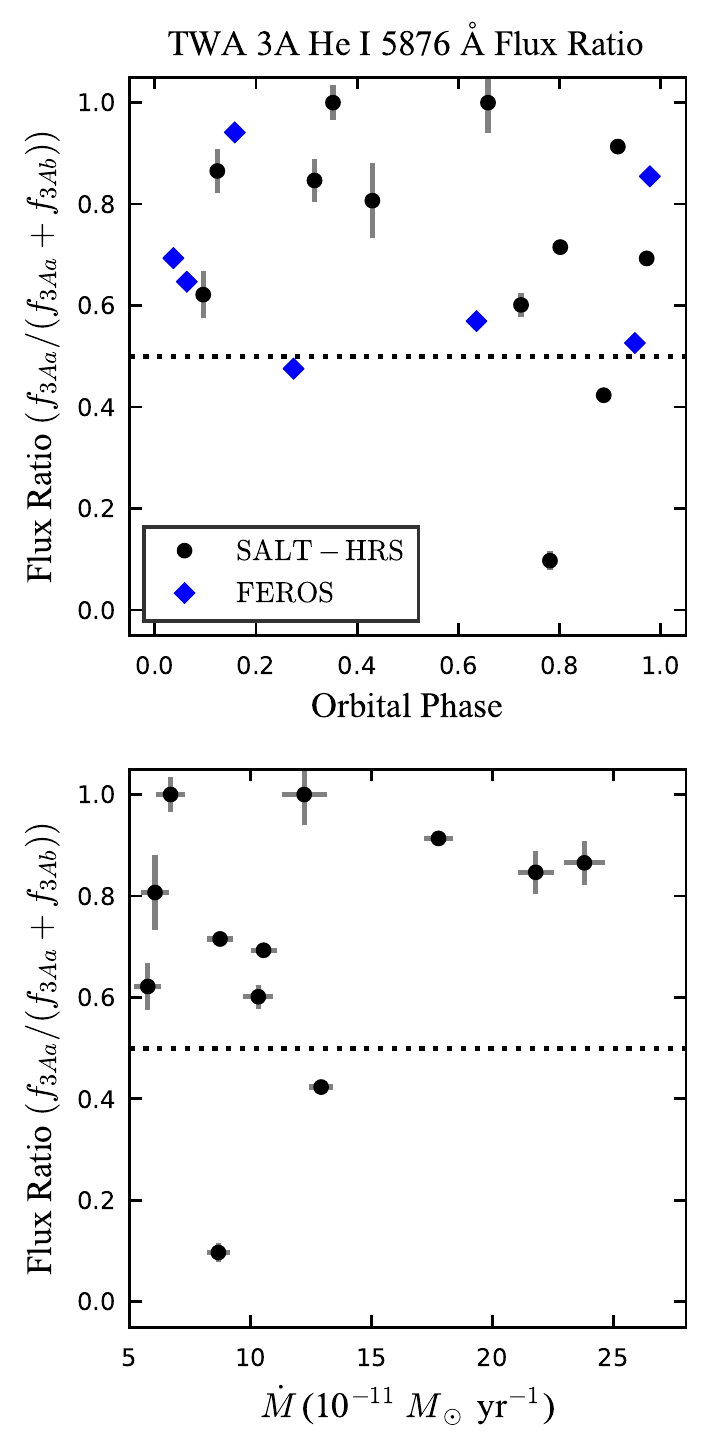}
  \caption{\hei \ accretion flux ratio for the TWA 3A binary, represented as the primary flux divided by the total flux. Flux ratio values are plotted against the binary orbital phase in the top panel, and the photometrically derived mass accretion rate in the bottom panel. Measurements from the SALT-HRS are displayed as black circles, FEROS observations are blue diamonds. A horizontal dotted line separates regions where the primary (above) and secondary (below) dominate the accretion flux. Secondary flux values have been scaled to account for its lower mass and radius compared to the primary. In a vast majority of cases, the primary star accretes at a higher rate.}
  \label{fig:hei_ratio}
\end{figure}

We then determine the flux from each star by integrating the spectrum, weighted by the relative strength of each fit profile, i.e.
\begin{equation}
    f_{\rm 3Aa} = \int f(v) \frac{p_{\rm 3Aa}(v)}{p_{\rm 3Aa}(v)+p_{\rm 3Ab}(v)} dv,
\end{equation}
\begin{equation}
    f_{\rm 3Ab} = \int f(v) \frac{p_{\rm 3Ab}(v)}{p_{\rm 3Aa}(v)+p_{\rm 3Ab}(v)} dv,
\end{equation}
where $f(v)$ is the accretion spectrum, and $p_{\rm 3Aa}$ and $p_{\rm 3Ab}$ are the best-fit, skewed Gaussian profiles for TWA 3Aa and TWA 3Ab, respectively. Uncertainties on these measurements are computed following the same procedure for our EW measurements (see Section \ref{ew_meas_corr}). This flux assignment scheme is flexible in that it does not require the emission at each epoch to necessarily follow a skewed Gaussian, and is more physically motivated than simple separation of flux by velocity.

To convert the relative fluxes of the two stars to a relative accretion rate, we take advantage of the tight correlation that exists between the \hei \ 5876\AA \ luminosity and the accretion luminosity \citep{Dahm2008,Herczeg&Hillenbrand2008,Fangetal2009,Alcalaetal2014}. Before directly interpreting the relative flux as a relative accretion rate, we account for the difference in stellar properties between the two stars. The mass accretion rate ($\dot{M}$), as determined by the accretion luminosity ($L_{\rm Acc}$), has the following form:
\begin{equation}
  \dot{M}\simeq \frac{L_{\rm Acc}R_{\star}}{GM_{\star}} \left(
  1-\frac{R_{\star}}{R_{\rm in}} \right)^{-1},
\label{eqn:MA_2}
\end{equation}
where $R_{\rm in}$ is the inner disk magnetospheric truncation radius, assumed to be $\sim$5$R_\star$. Given the differences in the TWA 3Aa and TWA 3Ab stellar parameters, if both stars were to accrete at the same rate, the primary would produce more emission by a factor of $(M_{\star,1}R_{\star,2})/(M_{\star,2}R_{\star,1})$, or nearly 10\%. As such, we scale the secondary flux measurement by this factor for a one-to-one comparison.

Figure \ref{fig:hei_ratio} presents the ratio of the primary flux to the total flux, as a function of orbital phase (top) and the photometrically derived mass-accretion rate (bottom). In nearly every case where a meaningful flux decomposition can be made, the primary star dominates the accretion emission, independent of the orbital phase or total mass accretion rate. We interpret this result as evidence for a consistently higher level of accretion onto the primary star. We discuss the context and implication of this result in Section \ref{prefacc}.

\begin{figure}[!tp]
  \centering
  \includegraphics[keepaspectratio=true,scale=1.1]{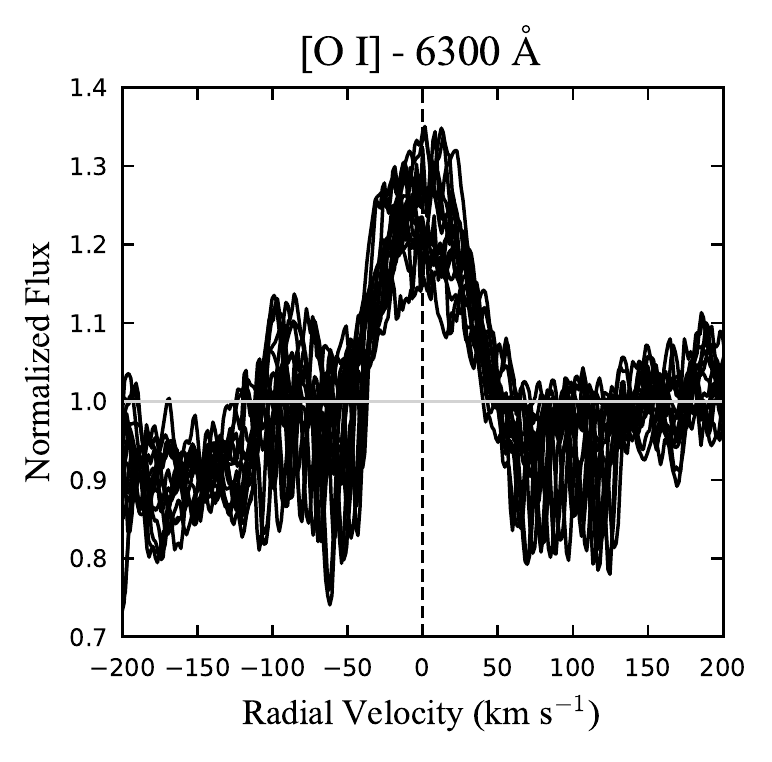}
  \caption{Velocity structure of \oi \ 6300\AA \ from SALT--HRS observations. The horizontal gray line marks the continuum level. Variations in the line profile amplitudes result from variable contribution of the TWA 3B continuum and are not reflective of intrinsic changes.}
  \label{fig:OI}
\end{figure}

\vspace{20pt}

\subsubsection{\oi \ 6300\AA}
\label{oi}

Forbidden oxygen emission is an important tracer of low-density gas in the outflows of young star-disk systems. In high accretion-rate systems \oi \ emission can be split into a high-velocity component which is spatially extended and a low-velocity component whose emission is unresolved near the central star \citep{Hartiganetal1995,Hirthetal1997}. It is generally accepted that the high-velocity components trace accretion-powered jet or micro-jet processes \citep{Rayetal2007}, while the low-velocity components trace a photoevaporative \citep{Ercolano&Owen2010,Ercolano&Owen2016} or magneto-centrifugal disk wind \citep[e.g.,][]{Baietal2016}. More recent studies of \oi\ at high spectral resolution find the low-velocity component can be further divided in to broad and narrow components whose kinematic behavior are most consistent with the MHD disk wind scenario \citep{Simonetal2016,Fangetal2018,McGinnisetal2018,Banzattietal2019}. In this interpretation, the broad component traces MHD disk winds at small disk radii ($\lesssim$0.5 AU), while the narrow components traces winds from more extended regions of the disk ($>$1 AU).

In Figure \ref{fig:OI} we present the velocity profiles of \oi \ 6300\AA. We find stable emission from a low-velocity
component that is invariant with changes in the stellar velocities and the mass accretion rate. Changes in profile amplitude result from variable continuum level from the triple companion (i.e., the corrected EW is constant across epochs). The average FWHM of these profiles is 54$\pm$4 \kms, in agreement with the single epoch analysed by \citet{Banzattietal2019}.

This kinematic behavior is typical of the low-velocity, narrow component, which \citet{Simonetal2016} find to be stable in single stars over decade timescales. The fact that only a low-velocity, narrow components is present in TWA 3A, which has likely carved out the central portion of its circumbinary disk at  0.3--0.5 AU scales, supports the interpretation that emission from this component originates at extended disk radii. The same \oi\ characteristics, namely stable, low-velocity emission, are also observed in the short-period, T Tauri binaries DQ Tau and KH 15D \citep{Huertaetal2005,Hamiltonetal2003}.

\section{Discussion}
\label{disc}

\subsection{The Binary Pulsed-Accretion Scenario}
\label{pulsed_acc}

Our analysis of the TWA 3A accretion-tracing emission lines reveals significant and consistent increases in the accretion rate near the binary periastron passage. The timing of these events is also consistent with more densely sampled, contemporaneous $U$-band photometry. With significant variation from orbit-to-orbit, we generally find very low accretion rates near apastron passage with accretion increasing near an orbital phase of $\sim$0.8, peaking near periastron, and quickly decaying by an orbital phase of $\sim$0.1. 

These results are in good agreement with simulations of binary accretion and provide strong support for a scenario in which streams of circumbinary material periodically feed the central binary. While the specific timing and morphology of the events observed in TWA 3A do not agree in detail with currently available simulations of accretion streams in eccentric binaries, the clear presence of near-periastron accretion events is encouraging.

A key issue in comparing current accretion-stream simulations to a short-period system like TWA 3A is the impact of stellar magnetic fields, which are not currently included in simulations. With a periastron separation of $\sim$20$R_\star$, the TWA 3A binary orbit should dynamically truncate the outer extent of a circumstellar disk at $\sim$0.2$a$, or $\sim$8$R_\star$ (the periastron extent of the Roche lobe; \citealt{Eggleton1983,Miranda&Lai2015}). This is similar to the inner magnetic truncation radius of circumstellar disks typically adopted for single stars ($\sim$5$R_\star$). The combination likely dramatically alters, if not entirely precludes, the existence of circumstellar disks in TWA 3A.

The presence or absence of stable circumstellar disk material (i.e., in bound Keplerian orbits) may fundamentally change the physical mechanisms which drive near-periastron accretion events. With a steady reservoir of circumstellar material periodically replenished by accretion streams, recent simulations suggest that accretion bursts are triggered by tidal torques that each star induces on its companion's disk \citep{Munoz&Lai2016}. Without circumstellar disks, accretion events depend on the delivery of circumbinary material directly to the stellar surface, or in practice a defined accretion radius \citep{Artymowicz&Lubow1996}. The simulation of both scenarios, however, produce similar predictions for the timing of accretion events. 

In actual PMS-binaries, accretion in the disk-less scenario is likely to depend sensitively on each star's detailed magnetic field strength and structure. The interaction of incoming stream material with stellar magnetic fields may determine the radius at which material is captured, and how it eventually reaches the stellar surface. This behavior could be different for each star.

While stable circumstellar disks seems unlikely in the TWA 3A system, our observations cannot explicitly rule them out. As our spectroscopic and photometric measurements trace material at the foot points of accretion flows, they do not describe the delivery of that material. The very low accretion rates near apastron are also consistent with both scenarios and do not provide a distinguishing lever. Observations directly probing the presence of circumstellar disks will be required for a definitive conclusion. 

\subsection{Preferential Accretion onto the TWA 3A Primary}
\label{prefacc}

From our analysis of the \hei \ 5876\AA \ velocity structure, we are able to conclude that the primary star of TWA 3A is consistently accreting at a higher rate than its binary companion during the epochs we have probed. We interpret this behavior as the result of circumbinary accretion flows that preferentially feed TWA 3Aa over its lower mass companion, TWA 3Ab. This conclusion is irrespective of the possible accretion scenarios discussed in Section \ref{pulsed_acc}, and places a strong observational constraint on the binary accretion-stream theory. As far as we are aware, this is the first observation of the relative accretion rates between the components of a PMS spectroscopic binary and is the first measurement of relative accretion rates in any binary that span multiple dynamical timescales. 

This result becomes more significant in the light of a large body of theoretical work that consistently predicts that the secondary should be the dominant recipient of circumbinary accretion flows \citep{Hayasakietal2007,Hayasakietal2013,Cuadraetal2009,Roedigetal2011,Farrisetal2014,Youngetal2015,Young&Clarke2015}. Secondaries are expected to dominate the mass accretion budget most at low mass ratios, with the relative accretion rate between the components approaching unity as does the mass ratio. These works have focused mainly on black-hole binary systems but should, in principle, describe the same physical processes present in PMS binaries. The physical interpretation of these simulations is that inner circumbinary disk material will have a specific angular momentum that is closer to that of the secondary than the primary, requiring less angular momentum loss in order to be accreted. Framed in another way, the secondary is always closer to the circumbinary disk allowing it to more easily collect material from an accretion stream. 

The degree to which the secondary accretes over the primary has been shown to depend on the disk ``temperature'', described as the ratio of the gas sound speed to the stellar orbital speed, $c = c_s / \sqrt{G(M_1+M_2)/a}$. For the range of disk temperatures simulated, $c$=0.05--0.25, \citet{Young&Clarke2015} find the hottest simulations approached a balance between the primary and secondary accretion rates, but were unable to consistently make the primary the dominant accretor in the system. Protoplanetry disks around low-mass stars will have disk temperatures near $c$$\sim$$0.05$ \citep{Youngetal2015}. For TWA 3A's mass ratio and a $c$=0.05 disk, these simulations would predict the primary should receive $\sim$40\% of the total accreted mass. On average, we find the TWA 3A primary accretes $\sim$70\% of the total mass.

The only recent simulations that have shown the possibility for primary dominated accretion are those of \citet{Dunhilletal2015} and \citet{Munoz&Lai2016}. These work stands out among those listed above by simulating thousands of binary orbital periods, thereby reaching dynamical relaxation in the circumbinary disk out to $\sim$10$a$. Once relaxed, individual accretion rate variability is tracked over many hundreds of binary orbital periods. This is in contrast to most modeling efforts that typically simulate the binary disk interaction for a few hundred orbital periods and track accretion rates for a few tens of orbits. Due to computational limitations most models are not able to capture secular variability. 

In the case of an eccentric (e=0.5), equal-mass binary, \citet{Munoz&Lai2016} find that the dominant recipient of circumbinary accretion flows alternates between the binary components  on timescales of hundreds of binary orbital periods. The driver of this behavior is the precession of an eccentric inner circumbinary disk cavity. This eccentricity naturally arises from the binary's non-axisymmetric potential. As this eccentric gap experiences apsidal precession, its orientation brings one star's apastron passage closer to the circumbinary disk than the other, inducing a larger accretion stream, and corresponding accretion burst compared to its companion. At any given time, one star may accrete 10-20 times more than its companion, but over many precession times, the integrated mass accreted by each component approaches unity. Although this result is for an equal-mass binary, the behavior should also take place to some degree at lower mass ratios. \citet{Dunhilletal2015} find a similar result with simulations tailored to the Herbig Ae binary system HD 104237 (DX Cha), which has an eccentricity of $e$=0.6 and a mass ratio of $q$=0.64. 

We note that some efforts to simulate specific T Tauri binary systems have found preferential accretion onto one component of an equal mass system (DQ Tau; \citealt{DeVal-Borroetal2011}), and preferential accretion onto the primary (rather than secondary) component under certain orbital parameters (GG Tau; \citealt{Gunther&Kley2002}). These specific results should be interpreted with caution, however, as they only simulate a small number of binary orbital periods. As a result, these studies may be depicting behavior that is transient in nature. 

If the precession of an inner circumbinary disk gap is indeed the source of our observed preferential accretion onto the TWA 3A primary, we should expect this behavior to reverse on half of the gap-precession timescale. For a pressureless particle disk, the precession rate around an eccentric binary is:
\begin{equation}
\dot{\omega_d}\simeq\frac{3\Omega}{4}\frac{q}{(1+q)^2}\left(1+\frac{3}{2}e^2\right)\left( \frac{a}{a_d}\right)^{7/2},
\end{equation}
where $\Omega$ is angular frequency of the binary, and $a_d$ is the inner edge of the circumbinary disk \citep{Munoz&Lai2016}. Assuming the dominant accretor flips after 180\degs \ of precession, the accretion flipping timescale then becomes $T_{\rm{flip}} \simeq \pi/\dot{\omega_d}$. Because this function has a steep dependence on $a_d$, a mass-weighted inner circumbinary disk radius is likely to provide the most representative value. \citet{Mirandaetal2017} find this value to vary from 3--10$a$ depending on the orbital configuration. For TWA 3A's orbital parameters, this corresponds to 80--5000 orbital periods for $a_d$ values of 3 and 10, respectively. Recent work by \citet{Leeetal2019}, however, has suggested that the precession of disk eccentricities is more dependent on the disk pressure profile and self gravity than the binary mass ratio and cavity size. And finally, \citet{Dunhilletal2015} measure an empirical $T_{\rm{flip}}$ value of 350 orbital periods in their HD 104237 tailored simulation.

With only two sets of observations spanning $\sim$150 orbital periods and no clear prediction for the inner-disk precession timescale, we have limited leverage to constrain the predicted accretion-flipping behavior. Continued observations of TWA 3A will be required to make a stronger statement about the validity of the circumbinary gap-precession scenario. Additionally, if the $T_{\rm{flip}}$ timescale is indeed many thousands of orbital periods, conducting similar observations on a sample of accreting binaries may provide a population that is capable of addressing the questions above. 

One last scenario we consider is the effect stellar magnetic fields have on the accretion process in short-period binaries. In the absence of stable circumstellar disks, dynamically driven flows from the circumbinary disk are likely to be mediated in some way by the magnetic fields before reaching the stellar surface. The details of this interaction will depend on the strength and configuration of the stellar magnetic field, which could be different for each component. If differences between the two stars affect the efficiency at which they capture or entrain incoming accretion flows, magnetic fields may play a comparable role to disk temperature and secular dynamics in determining which star is the dominant accretor in a binary. 

In a broader context, the results presented here have important implications for binary evolution. For instance, circumbinary accretion is often employed to explain the abundance of ``stellar twins'' \citep[$q>0.95$,][]{Raghavanetal2010,Moe&DiStefano2017}, assuming accretion dominated by the secondary equalizes the mass ratio. Given the discussion above, however, we hesitate to draw any broad conclusions. It is unclear the degree to which the TWA 3A accretion behavior is representative of other T Tauri binaries or to proto-binary systems where, due to higher disk masses and mass accretion rates, the largest change in the binary mass ratio is likely to take place. We advocate for continued examination of the above assumption and for future theoretical efforts to examine secular interaction between binaries and their disks and the effects of stellar magnetic fields.

\section{Summary}
\label{conc}

In this work we have monitored the T Tauri binary TWA 3A with time-series, high-spectral-resolution optical spectroscopy and time-series optical photometry to trace variability in the timing, amplitude, and kinematics of accretion events. The main results of our work are as follows: 
\begin{enumerate}

\item Accretion-tracing emission lines H$\alpha$, H$\beta$, and \hei \ 5876\AA \ display increased EWs, line widths, and line structures near periastron passage, indicative of enhanced accretion. The timing of these events mirrors the behavior of spectroscopic veiling measurements, and the photometric accretion diagnostics presented in \citet{Tofflemireetal2017b}.

\item The strength and line shape of \hal \ outside of periastron appears to be dominated by chromospheric emission, suggesting very low accretion outside of these discrete periastron bursts.

\item The presence of orbital-phase-dependent accretion events, and specifically periastron accretion bursts, is in good agreement with simulations of eccentric binary accretion. Our results generally support a binary accretion scenario in which circumbinary accretion streams periodically feed the central binary. 

\item The velocity profiles of narrow, accretion-tracing emission lines are consistently found to center on the radial velocity of the primary star in TWA 3A. We have focused our analysis most heavily on \hei \ 5876\AA, but note the same behavior is present to some degree in H$\gamma$, H$\delta$, \hei \ 4471\AA, and \caii \ H and K. 

\item To enable a detailed analysis of \hei \ 5876\AA \ emission, we develop a spectral decomposition scheme utilizing spectral line broadening functions to remove the contributions from the binary's triple companion, veiling, and the chromospheric components of all three stars. Left with emission from accretion alone, we separate the flux from both stars, finding the TWA 3A primary typically contributes $\sim$70\% of the accretion emission (after correcting for the difference in their stellar parameters). These are the first measurements of their kind for a spectroscopic binary. 

\item We interpret the primary-dominated \hei \ 5876\AA \ emission as evidence that circumbinary accretion streams are preferentially feeding the TWA 3A primary. This result is in conflict with the vast majority of numerical simulations of binary accretion that suggest the secondary should be the dominant accretor. The simulations of \citet{Munoz&Lai2016}, which are able to capture secular evolution, develop a precessing, eccentric inner disk cavity that allows for the dominant accretor to alternate between the primary and secondary on timescales of hundreds of orbital periods. If this is the case, future observations should find \hei \ 5876\AA \ emission to preferentially occur at the secondary's radial velocity near periastron passage. 

\item The emission of \oi \ 6300\AA \ consists of a single, low-velocity component that is not correlated with the accretion rate or the stellar radial velocities, favoring a disk wind scenario originating in the extended circumbinary disk. 
\end{enumerate}

\acknowledgments 
We would like to thank the following people: Guillermo Torres and Eike Guenther for providing archival FEROS spectra of TWA 3, Diego Mu{\~n}oz and Enrico Ragusa for useful discussions of circumbinary accretion theory, Adam Kowalski and John Wisniewski for discussions of chromospheric emission, Adam Kraus for discussions on broadening function applications, Sebastian Heinz for his comments on the thesis version of this manuscript, Greg Herczeg for his general young-star expertise, and Ken Nordsieck for his help in designing our SALT program. We like to acknowledge the entire SALT observing team, especially Encarni Romero Colmenero, Petri Vaisanen, Steven Crawford, Luke Tyas, Brent Miszalski, Alexei Kniazev, and Paul Kotze.  Support for this research was provided by the University of Wisconsin-Madison, Office of the Vice Chancellor for Research and Graduate Education with funding from the Wisconsin Alumni Research Foundation.

\facilities
Some of the observations reported in this paper were obtained with the Southern African Large Telescope (SALT). This work makes use of observations from the LCO network. Part of this work is based on observations made with ESO Telescopes at the La Silla Paranal Observatory under programme ID 086.C-0173. The authors acknowledge the Texas Advanced Computing Center (TACC) at The University of Texas at Austin for providing HPC resources that have contributed to the research results reported within this paper (\url{http://www.tacc.utexas.edu}). This work has made use of data from the European Space Agency (ESA) mission {\it Gaia} (\url{https://www.cosmos.esa.int/gaia}), processed by the {\it Gaia} Data Processing and Analysis Consortium (DPAC, \url{https://www.cosmos.esa.int/web/gaia/dpac/consortium}). Funding for the DPAC has been provided by national institutions, in particular the institutions participating in the {\it Gaia} Multilateral Agreement. This research has made use of NASA's Astrophysics Data System.

\software
{The spectral-line broadening function code used in this project is available from \url{https://github.com/tofflemire/saphires} \citep{saphires}. This research has also utilized matplotlib \citep{Hunter2007}, scipy \citep{scipy}, numpy \citep{numpy}, and astropy \citep{astropy2,astropy1} packages. STSDAS and PyRAF are products of the Space Telescope Science Institute, which is operated by AURA for NASA.}


\begin{appendices}

\section{SALT--HRS Red-Arm Specific Reductions}
\label{ap:red}

\begin{figure}[!b]
  \centering
  \includegraphics[keepaspectratio=true,scale=0.9]{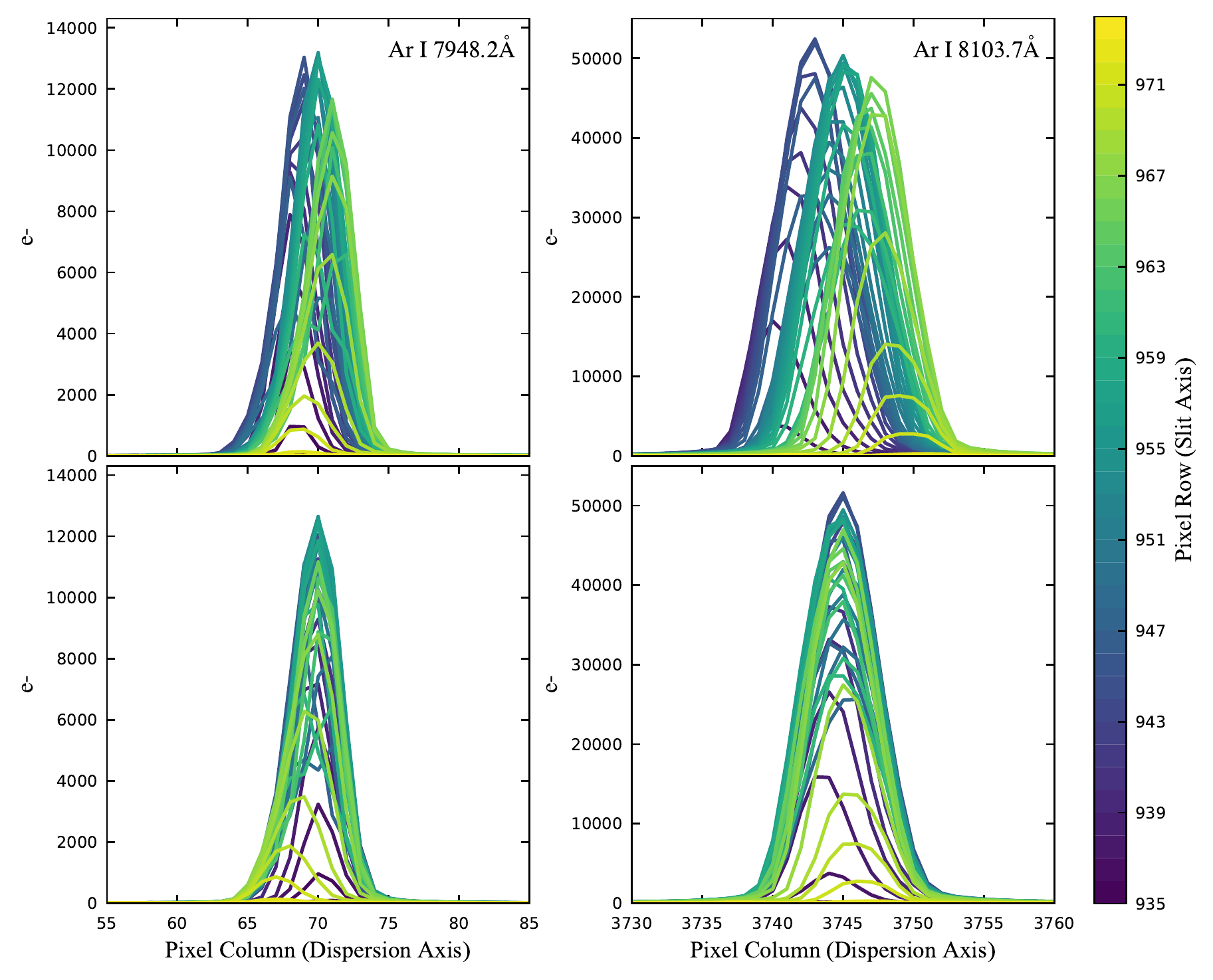}
  \caption{Two \ari \ emission lines, 7948.2\AA \ and 8103.7\AA, are presented on the left and right columns, respectively, extracted from the 62nd spectral order ($\sim$7945\AA \ to 8110\AA) of a Th-Ar comparison in the SALT--HRS red arm. All four panels display the distribution of the emission line's flux across the dispersion axis (an image of the pseudo-slit). The line color signifies the pixel row across the vertical extend of the spectral order. Flux profiles prior to pseudo-slit rectification are shown in the top panels with the transformed version below. (Pixel rows displayed here are extracted from the order rectified image.)}
  \label{fig:redarm}
\end{figure}

As described in Section \ref{redarm}, geometric distortions of the HRS pseudo-slit are variable across the dispersion axis in red arm spectral orders. This results in an increasingly broad resolution element with wavelength in a given spectral order, which distorts the shapes of lines when combining overlapping spectral orders. The main culprit of this distortion is a tilt in the pseudo-slit image that increases with pixel column (although we note that the even without the tilt, the pseudo-slit's FWHM in a given pixel row also increases with pixel column). 

The top panels of Figure \ref{fig:redarm} display this distortion with two emission lines from the same spectral order of a Th-Ar comparison spectrum (m=62). The left panel shows an emission line, \ari \ 7948.2\AA, which falls on the left side of the CCD. Each curve is the emission profile of this line for an individual pixel row in the spectral order. The line color corresponds to the pixel row which is presented in the color bar at the right. Here it is clear that the line peak falls at smaller column-pixel values at the bottom of the spectral order and higher column-pixel values at the top. On the right side of the CCD, this behavior is more extreme where from the top of the spectral order to the bottom, the pseudo-slit image has shifted by more than the line profile's FWHM value. One Ar line that is severely affected in this way is presented in the top right panel of Figure \ref{fig:redarm}. Simply summing the flux across pixel-columns, as most extraction methods do, would smear the line's flux, extensively lowering the spectral resolution. 

Our scheme to correct this distortion exploits the fact that emission lines in a Th-Ar comparison spectra provide an image of the pseudo-slit at discrete wavelengths across the spectral order. By measuring the ``ridge line'' of many emission lines in a spectral order, we are able to map the tilt of the pseudo-slit as a function of the column pixels and row pixels within the spectral order. This mapping can then be used to transform, or ``rectify'' the pseudo-slit.

The correction of this distortion, as we have conceived it here, requires spectral orders that fall directly across pixel rows, with no curvature across the CCD. This is not case for raw HRS red arm images, nor is it for any echelle spectrograph. As such, we first rectify the spectral orders using a spectral flat with the IRAF, NOAO.longslit functions fitcoords and transform. (For reference, the spectral lines presented in top panels of Figure \ref{fig:redarm} are extracted from single pixel rows of the order rectified image.)

For each spectral order the center row is set as the zero-point about which pixel rows above and below are transformed to match. For each pixel row we record the location of emission line's peaks with a Gaussian fit. The column-pixel locations of these line peaks, $x_{\rm{row}}$, are then fit to the peak at the order center, $x_{\rm{center}}$, with the following expression:
\begin{equation}
  x_{\rm{center}} = x_{\rm{row}} + Ay_{\rm{offset}}x_{\rm{row}} + By_{\rm{offset}}x_{\rm{row}}^2 + Cy_{\rm{offset}},
\label{eqn:red_trans}
\end{equation}
where $y_{\rm{offset}} = y-y_{\rm{center}}$, the number of pixel rows away from the order center. The best-fit values for $A$, $B$, and $C$ then define the correction for a given row where Equation \ref{eqn:red_trans} is applied as a geometric coordinate transformation. In order to conserve flux during this process, we multiply the transformed flux values by the determinant of the transformation's Jacobian matrix. We find this preserves the flux values of each row to within 1\%. 

The bottom panels of Figure \ref{fig:redarm} presents the pseudo-slit-rectified line profiles where the pseudo-slit is aligned with the pixel columns, signified by their overlapping profiles. This process reduces the size of the resolution element at all points across the order and increases the consistency across wavelength. 

We note that the pseudo-slit image width increases for every pixel row with increasing pixel column. This can be seen as a wider distribution in the bottom right panel compared to the bottom left. This behavior is present prior to order rectification and is not introduced by it. The effect is minor, however, compared to the pseudo-slit tilt and we do not attempt to correct for it. 

\section{Broadening Function Template Selection}
\label{ap:bfs}

\begin{figure}[!b]
  \centering
  \includegraphics[keepaspectratio=true,scale=0.47]{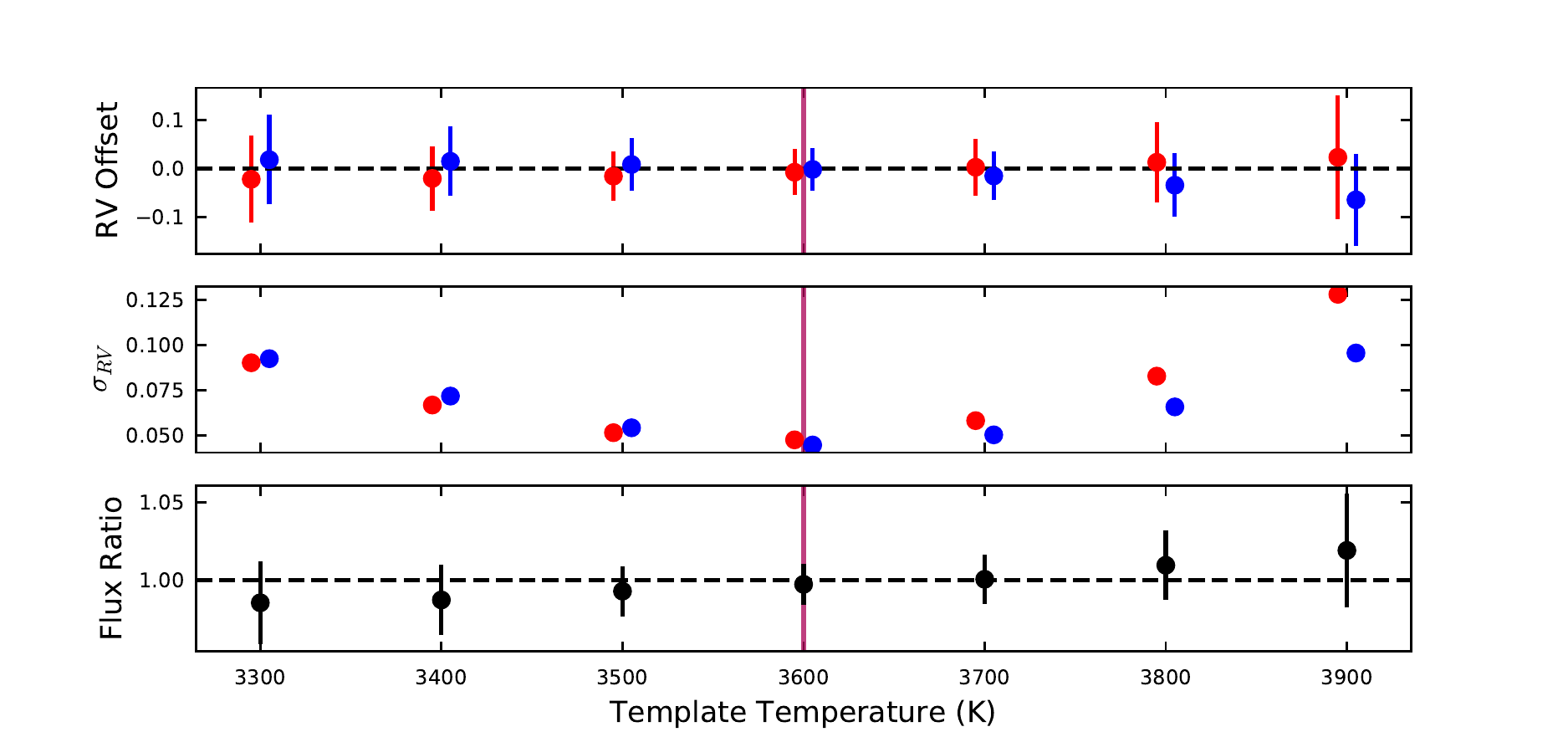}
  \includegraphics[keepaspectratio=true,scale=0.47]{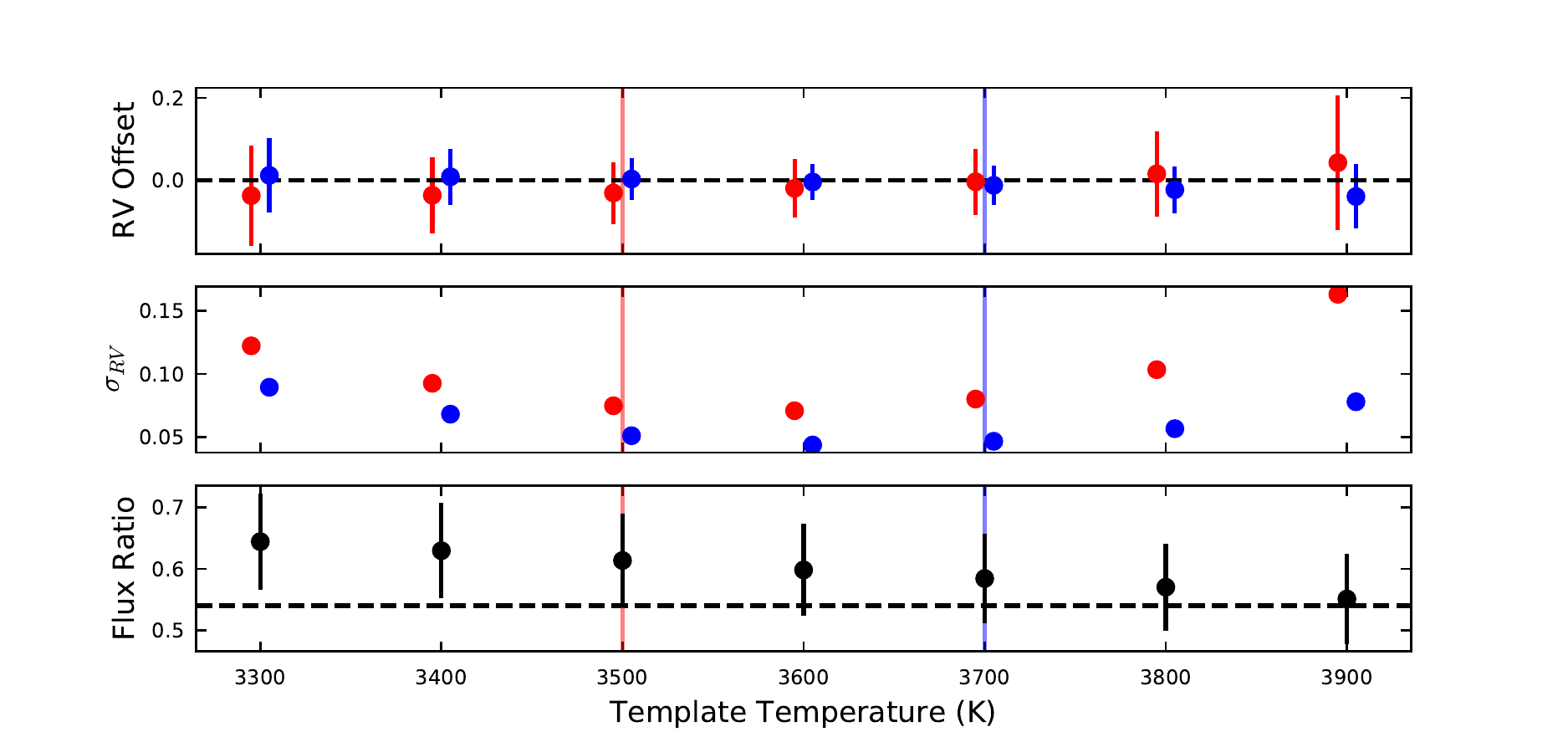}
  \includegraphics[keepaspectratio=true,scale=0.47]{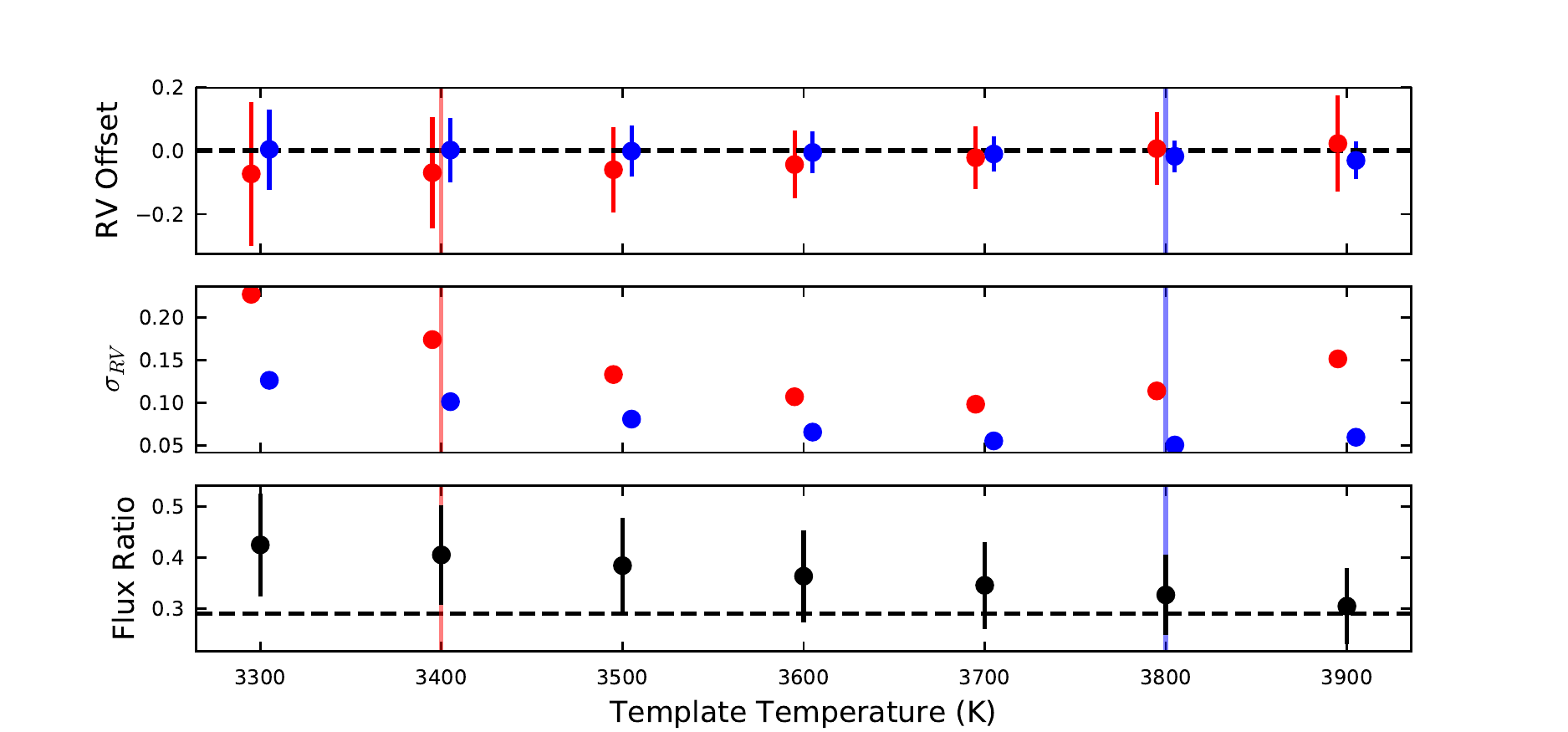}
  \caption{Input-recovery results for synthetic spectroscopic binaries made from BT-Settl model spectra \citep{Baraffeetal2015}. Each group of three panels presents the measured RV offset from the input ($\pm$20 km s$^{-1}$), the standard deviation in the RV measurement from order-to-order, and the measured flux ratio as a function of the BF template.  Blue and red points are measurements for the primary and secondary, respectively. Error bars represent the standard deviation of measurements from order-to-order. Vertical blue and red lines mark the spectra used to make the synthetic binary. Horizontal dashed lines mark the input values.}
  \label{fig:bf_test}
\end{figure}

In order to develop a criterion for selecting the best template for our BF analysis, we perform an input-recovery analysis with synthetic spectroscopic binaries created from BT-Settl model spectra \citep{Baraffeetal2015}. We create three synthetic binaries with different temperature components, namely, 3600--3600 K (Figure \ref{fig:bf_test}; left), 3700--3500 K (Figure \ref{fig:bf_test}; right), and 3800--3400 K (Figure \ref{fig:bf_test}; bottom). The primary (hotter) component is shifted to $-$20 \kms \ and the secondary is shifted to +20 \kms. The spectra are smoothed to a spectral resolution of $R$$\sim$40,000 and noise is added to mimic a S/N of 50. Our analysis focuses on the wavelength range 5000--7000\AA.

We compute the BF for templates spanning temperatures 3900 to 3400 K, breaking up the spectra into 20, 100\AA \ orders. Figure \ref{fig:bf_test}, presents our measurements for each synthetic binary as a function of the BF templates. In each set of panels, the recovered RV, the RV uncertainty, and flux ratio are presented from top to bottom. Blue and red points mark the primary and secondary, respectively, and are slightly offset to the right and left to avoid overlapping points. Uncertainties in each measurement are determined from the standard deviation from order-to-order. The spectral templates used to create the synthetic binary are represented as vertical blue (primary) and red (secondary) lines, and horizontal dashed lines mark the RV and flux ratio input value. 

We find that the measured RVs are always consistent with the input values within 1--$\sigma$, but errors are smallest for templates similar to the input primary. The template that produces the smallest uncertainty in the primary's RV also recovers the flux ratio within 1--$\sigma$. Given this result, we select templates from our empirical samples that produce the smallest uncertainty in the primary's RV. 

\section{Additional Spectral Line Variability}
\label{ap:lines}

\begin{figure}[!b]
  \centering
  \includegraphics[keepaspectratio=true,scale=0.9]{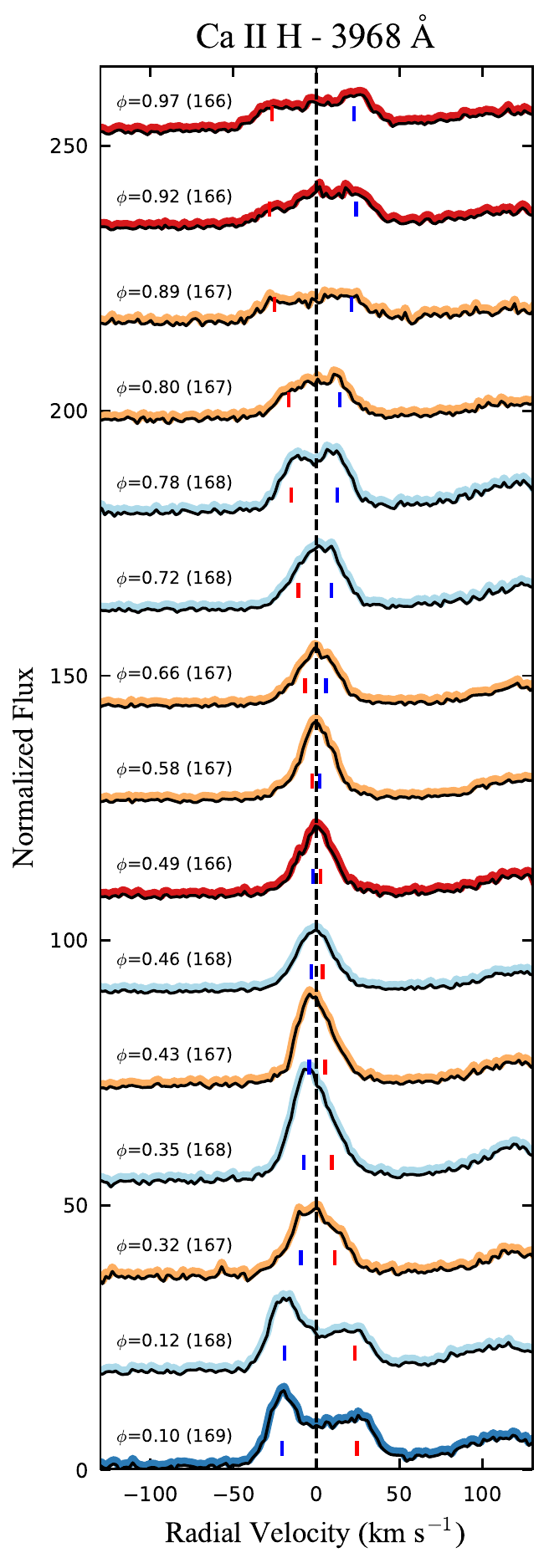}
  \includegraphics[keepaspectratio=true,scale=0.9]{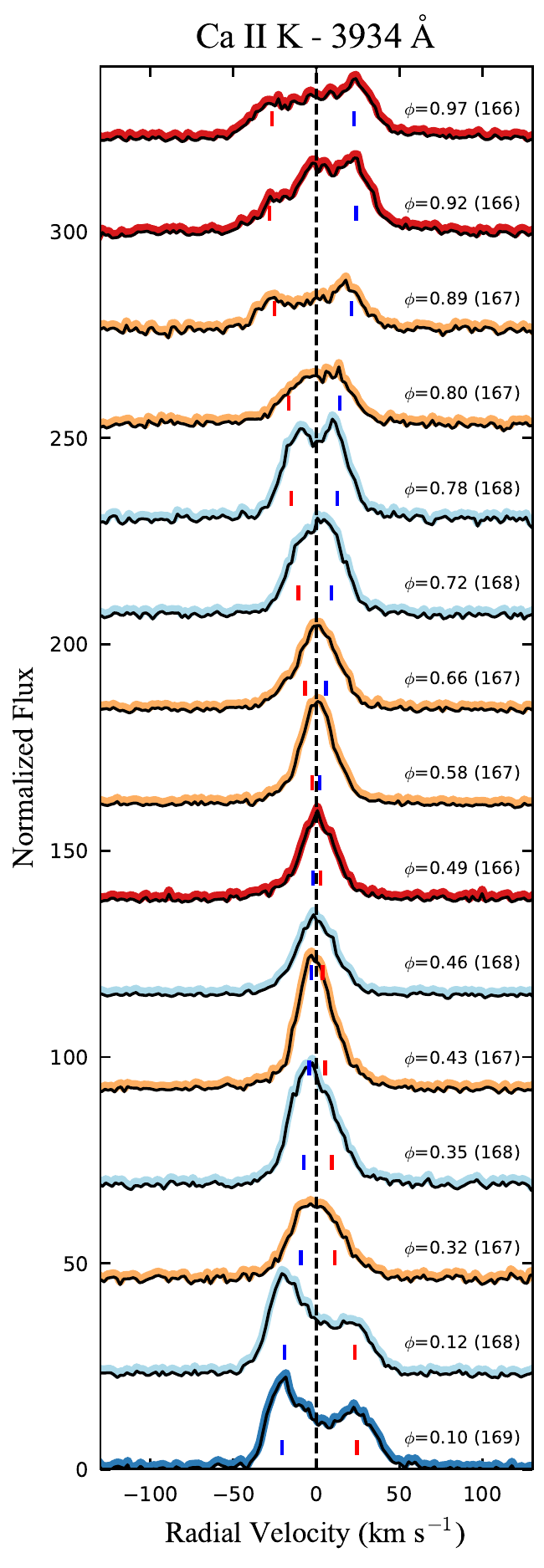}
  \includegraphics[keepaspectratio=true,scale=0.9]{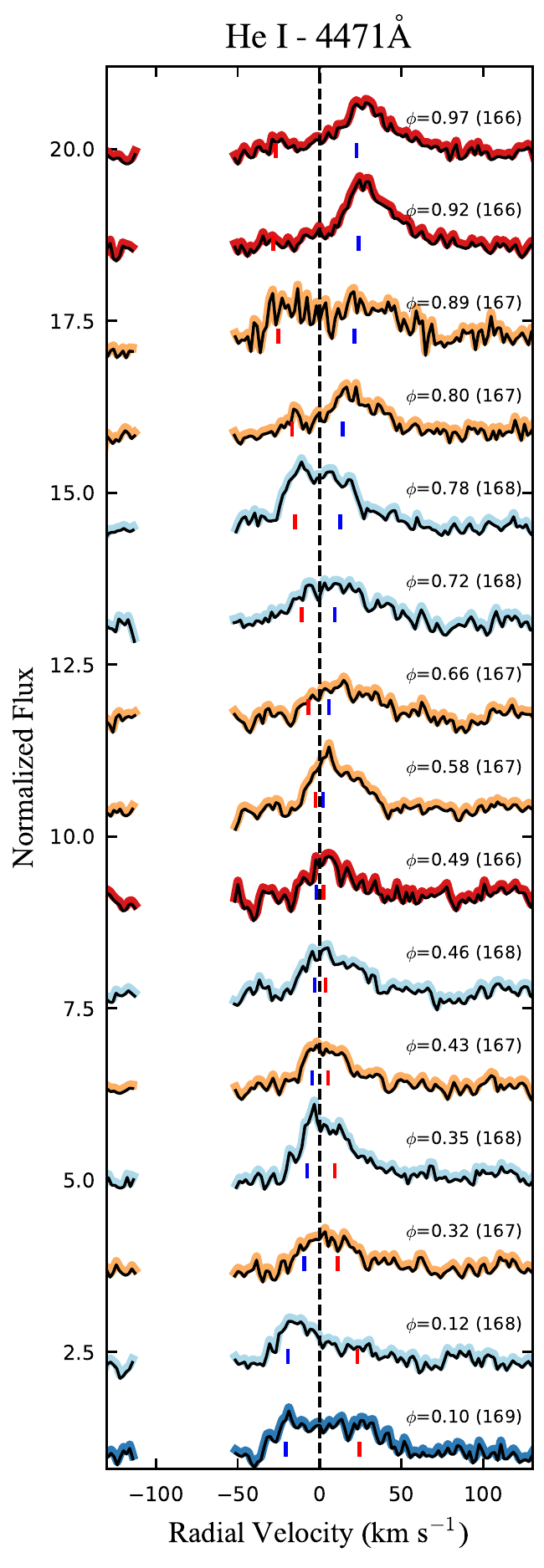}
  \caption{Velocity structure of \caii \ H and K and \hei \ 4471\AA \ emission lines from SALT--HRS observations. Rising flux values at positive velocities in the \caii \ H panel is the contribution of H $\epsilon$. For \caii \ H and K panels, the stellar continuum is not consistently detected at these wavelengths; normalization is with respect to the sky/detection background. Only lines shapes, not their amplitude, are comparable between these spectra. The gap at negative velocities in the \hei \ 4471\AA \ panel corresponds to a chip defect. (Figure has the same layout as Figure \ref{fig:balmer}, see a full description there.)}
  \vspace{5pt}
  \label{fig:cahkh}
\end{figure}

Figure \ref{fig:cahkh} presents the center-of-mass velocity structure for three accretion-tracing emission lines, \caii \ H and K, and \hei \ 4471\AA. The \caii \ line are excluded from our main analysis due to low levels of stellar continua at these wavelengths (see Section \ref{salt}), and the strong chromospheric component of these lines. \hei \ 4471\AA \ is also excluded due to its weak emission. All three lines are included in this appendix to highlight a similar behavior to \hei \ 5876\AA \ where the dominant component of emission is seen to follow the primary stellar velocity. 

\end{appendices}

\end{document}